\documentclass[a4paper]{article}
\usepackage{anysize,times}
\usepackage{rsflash}
\usepackage{psfig}
\def\I{\rm {\scriptsize I}}
\marginsize{0.5in}{0.5in}{0.5in}{0.5in}

\begin{document}
\date{  }
%
%

\title{The role of ionization in symbiotic binaries}

\author{Augustin Skopal \\
        Astronomical Institute, Slovak Academy of Sciences, 
        SK-059\,60 Tatransk\'{a} Lomnica, Slovakia} 

\maketitle

\vspace*{2cm}
{\large
Abstract 
\begin{enumerate}
\item Introduction
\item Ionization during quiescent phases \\
\hspace*{0.2cm}2.1 Observations \\
\hspace*{0.2cm}2.2 On the nature of the wave-like variation \\
\hspace*{1cm}2.2.1 A reflection effect \\
\hspace*{1cm}2.2.2 An ionization model and the vawe-like variability
\item Ionization during transition periods \\
\hspace*{0.2cm}3.1 Systematic variation in the O-C residuals \\
\hspace*{1cm}3.1.1 Common properties \\
\hspace*{0.2cm}3.2 Principle of apparent orbital changes \\
\hspace*{1cm}3.2.1 Asymmetric shape of the H\,\I\I\ zone \\
\hspace*{1cm}3.2.2 The cases of BF\,Cyg and AG\,Peg
\item Ionization during active phases \\
\hspace*{0.2cm}4.1 Eclipsing system CI\,Cyg \\
\hspace*{0.2cm}4.2 Non-eclipsing system AG\,Dra \\
\hspace*{0.2cm}4.3 A mechanism of outbursts \\
\hspace*{0.2cm}4.4 Mass loss in the symbiotic system CH\,Cyg \\
\hspace*{1cm}4.4.1 Mass loss from the H${\alpha}$ emission 
\item Conclusions
\end{enumerate}

Acknowledgments \\

References
}

\clearpage

\twocolumn


\begin{abstract}
Processes of ionization and recombination often influence signi\-ficantly 
the observed spectrum produced by symbiotic stars in a very wide 
wavelength region. 
Properties of the nebular spectrum provide important information 
on some fundamental parameters and on geometric structure of 
symbiotic stars. 
This paper reviews the recently recognized effects of
ionization/recombination in these binaries and summarizes first
results. 
We concentrate in this review on the continuum emission in the range 
from ultraviolet to near-IR wavelengths. 
\end{abstract}

\section{Introduction}

The objects, which are now commonly named as {\em symbiotic stars} were 
discovered at the beginning of the 20th century as stars with peculiar 
and/or combined spectra [13,30], 
because of the simultaneous presence of spectral features indicating 
two very distinct temperature regimes. Their present denotation was 
used for the first time by [29]. 
Currently the symbiotic stars are understood as interacting binary systems 
consisting of a cool giant and a hot compact star, which is in most cases 
a white dwarf. Typical orbital periods are between 1 and 3 years, but can 
be significantly larger. There are two principal processes of interaction 
in such long-period binaries: 

(i) Mass loss from the cool component -- assumed to be in the form of
a wind -- represents the primary condition for appearance of the symbiotic 
phenomenon. 

(ii) Accretion of a part of the material lost by the giant by its 
compact companion. This process generates a very hot 
($T_{\rm h} \approx 10^5$\,K) 
and luminous 
($L_{\rm h} \approx 10^2 - 10^4\,\rm L_{\odot}$)
source of radiation. 

On the basis of the way, in which the generated energy is being liberated,
we distinguish two phases of symbiotic binary: \\
A {\em quiescent phase}, during which the hot component releases its 
energy approximately at a constant rate and spectral distribution. 
The hot radiation ionizes a fraction of the neutral giant's wind, but 
can also be Raman and Rayleigh scattered by its neutral particles. 
The process of ionization gives rise to nebular 
emission comprising numerous lines of high excitation/ionization and 
the continuum. 
As a result the spectrum of symbiotic stars during quiescence consists 
of basically three components of radiation -- two stellar and one nebular. 
Figure 1 shows example of such the spectral energy distribution (SED) 
for AG\,Peg. \\
An {\em active phase}, during which the hot component radiation 
chan\-ges significantly, at least in its spectral distribution. 
In cases of a nova-like outbursts the luminosity rapidly increases on 
the time-scale of weeks/months, followed by its gradual decrease to 
the quiescent level within a few years or decades. A common feature 
of active phases is a high-velocity mass ejection. 

The aim of this contribution is to review the recent research results 
concerning to effects of ionization in symbiotic binaries, which 
suggest new insight to the nature of some well known phenomena 
observed in these objects. 
Section 2 deals with an orbitally-related variation in the light curves 
          during quiescent phases, 
section 3 summarizes the effect of apparent changes of orbital period 
          during transition from active to quiescent phases, and 
section 4 discuss some aspects of ionization during active phases. 
%
\begin{figure}   
  \centering     
  \centerline{
  \hbox{\psfig{figure=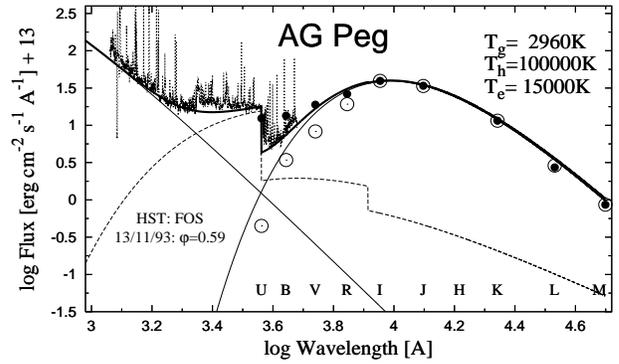,width=8cm,angle=-90}}}
\caption[]{
Reconstructed SED in the continuum of AG\,Peg between 0.12 and 5$\mu$m 
by a three component model of radiation. 
Solid thin lines are Planck functions, which match the continuum of 
the cool and the hot star, respectively. Dotted line represents the 
hydrogen f-b and f-f continuum. The solid thick line is the resulting 
continuum. Observations were dereddened with $E_{\rm B-V} = 0.1$. 
More about the fitting procedure is in [49]. 
}  
\end{figure}

\section{Ionization during quiescent phases}

In this section we discuss the orbitally-related variation observed 
in the optical/near-UV continuum within a simple ionization model 
of symbiotic binaries. 

\subsection{Observations}

The most pronounced feature observed in the light curves of symbiotic 
stars during phases of quiescence is a periodic wave-like variation as 
a function of the orbital phase. 
This type of variability was revealed by long-term multicolour 
photometry made in 1960's [1,2] for AG\,Peg 
and Z\,And. At present, all about 20 symbiotic objects, which have 
sufficiently covered light curves, display signatures of the wave-like 
variability during their quiescent phases. 
It is characterized by a large amplitude $\Delta m \sim$\,1\,mag or more,
which is a function of the wavelength -- we always observe 
$\Delta U > \Delta B \ge \Delta V$. The period is approximately equal
to the orbital period, and a minimum occurs at/around the inferior
conjunction of the cool component. These properties relate this type
of variations to the orbital motion. Figure 2 shows examples of
V1329\,Cyg, AG\,Peg, AG\,Dra, AX\,Per, BF\,Cyg and Z\,And.
%
\begin{figure*}
  \centering
  \centerline{\hbox{
  \psfig{figure=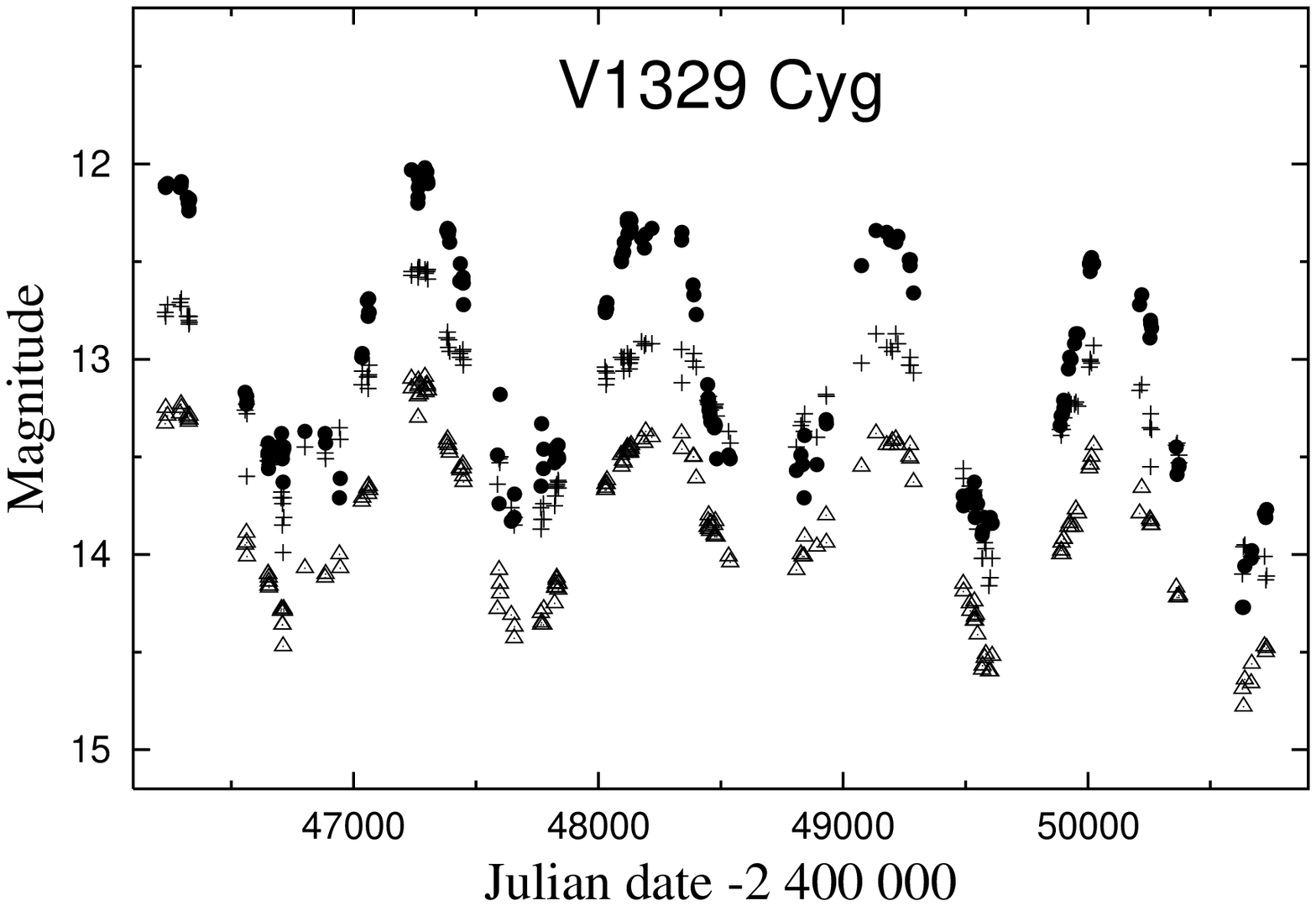,width=6cm,angle=-90}
\hspace*{0.1cm}
  \psfig{figure=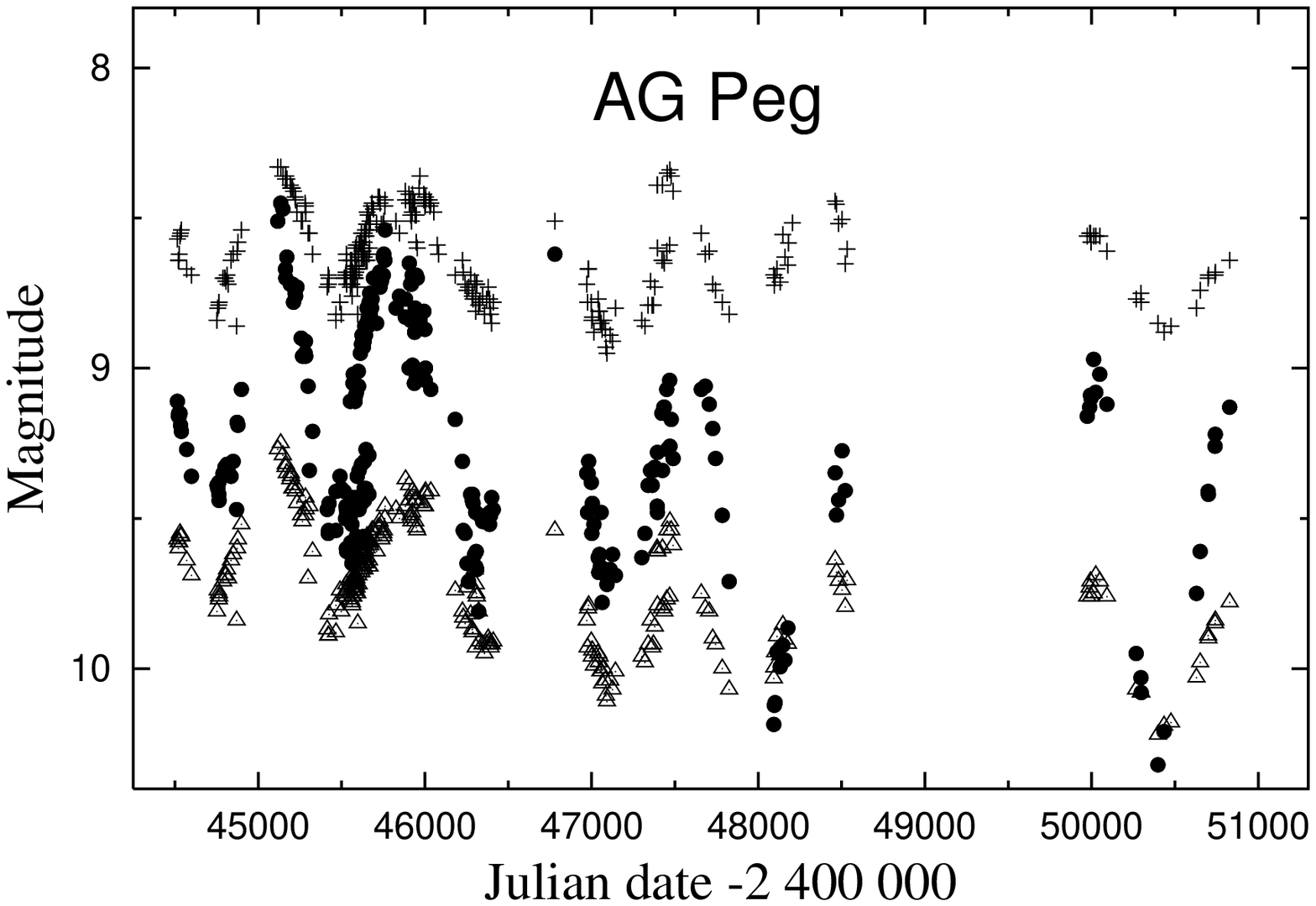,width=6cm,angle=-90}
\hspace*{0.1cm}
  \psfig{figure=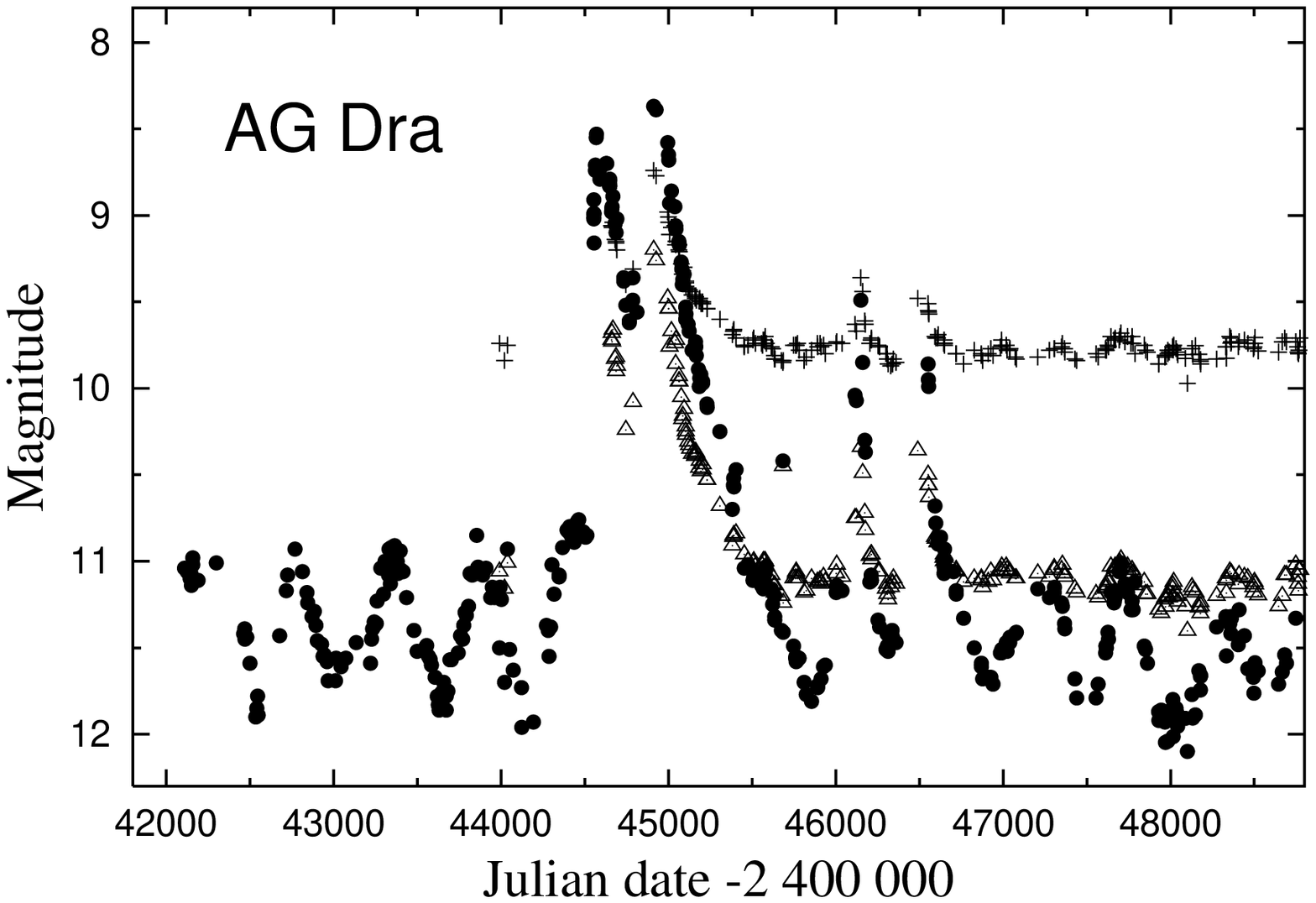,width=6cm,angle=-90}}}
\vspace*{0.1cm}
  \centerline{\hbox{
  \psfig{figure=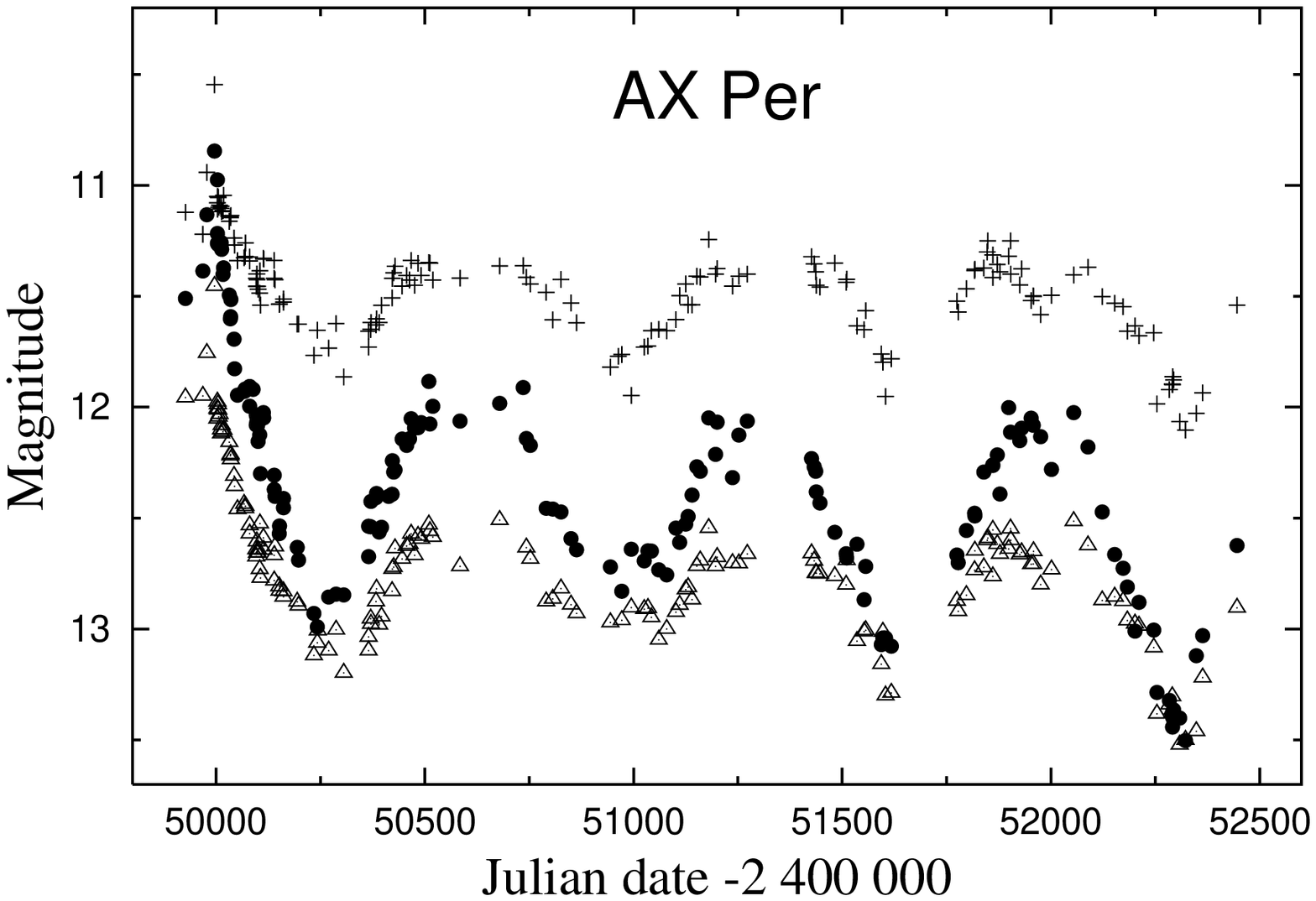,width=6cm,angle=-90}
\hspace*{0.1cm}
  \psfig{figure=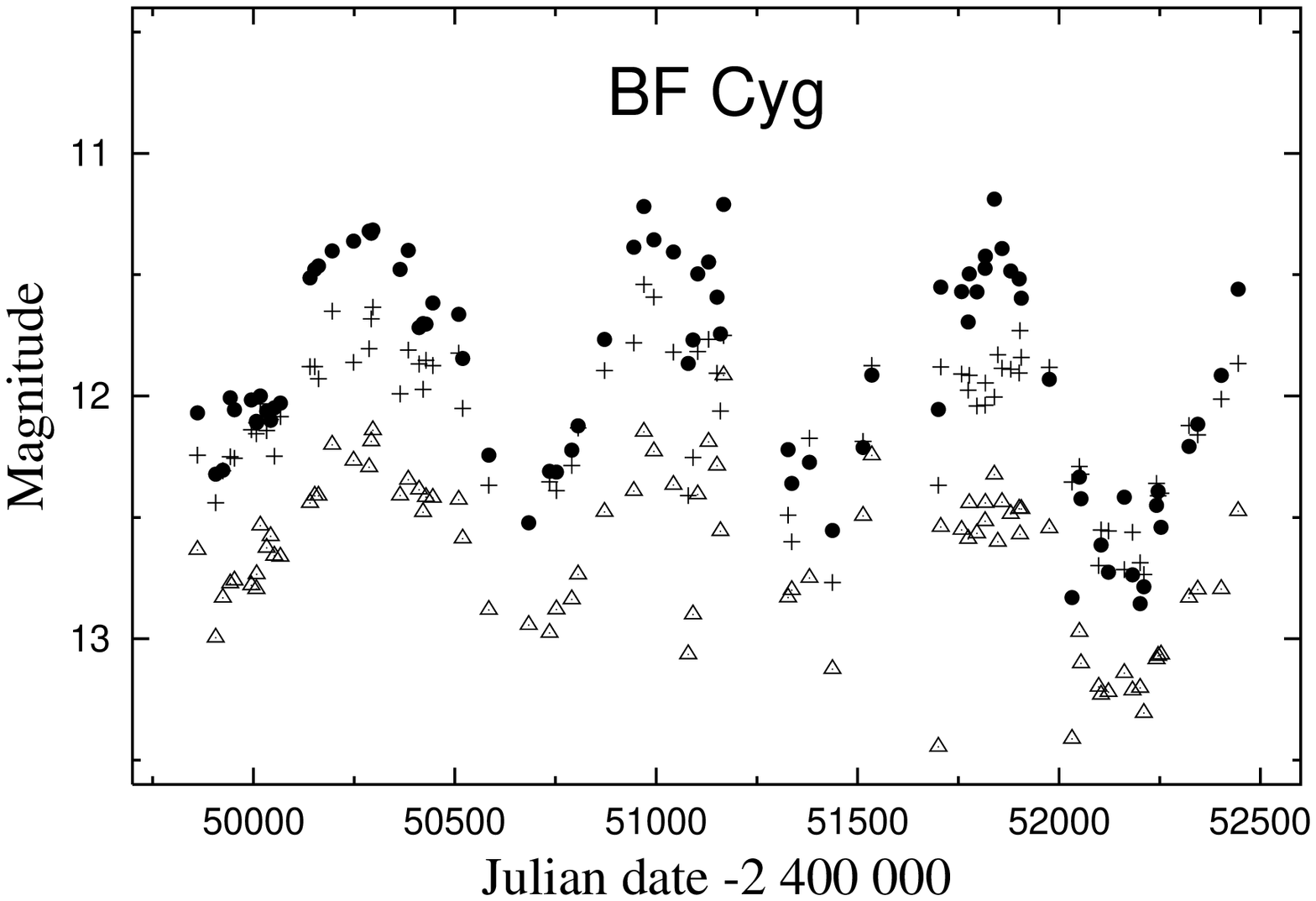,width=6cm,angle=-90}
\hspace*{0.1cm}
  \psfig{figure=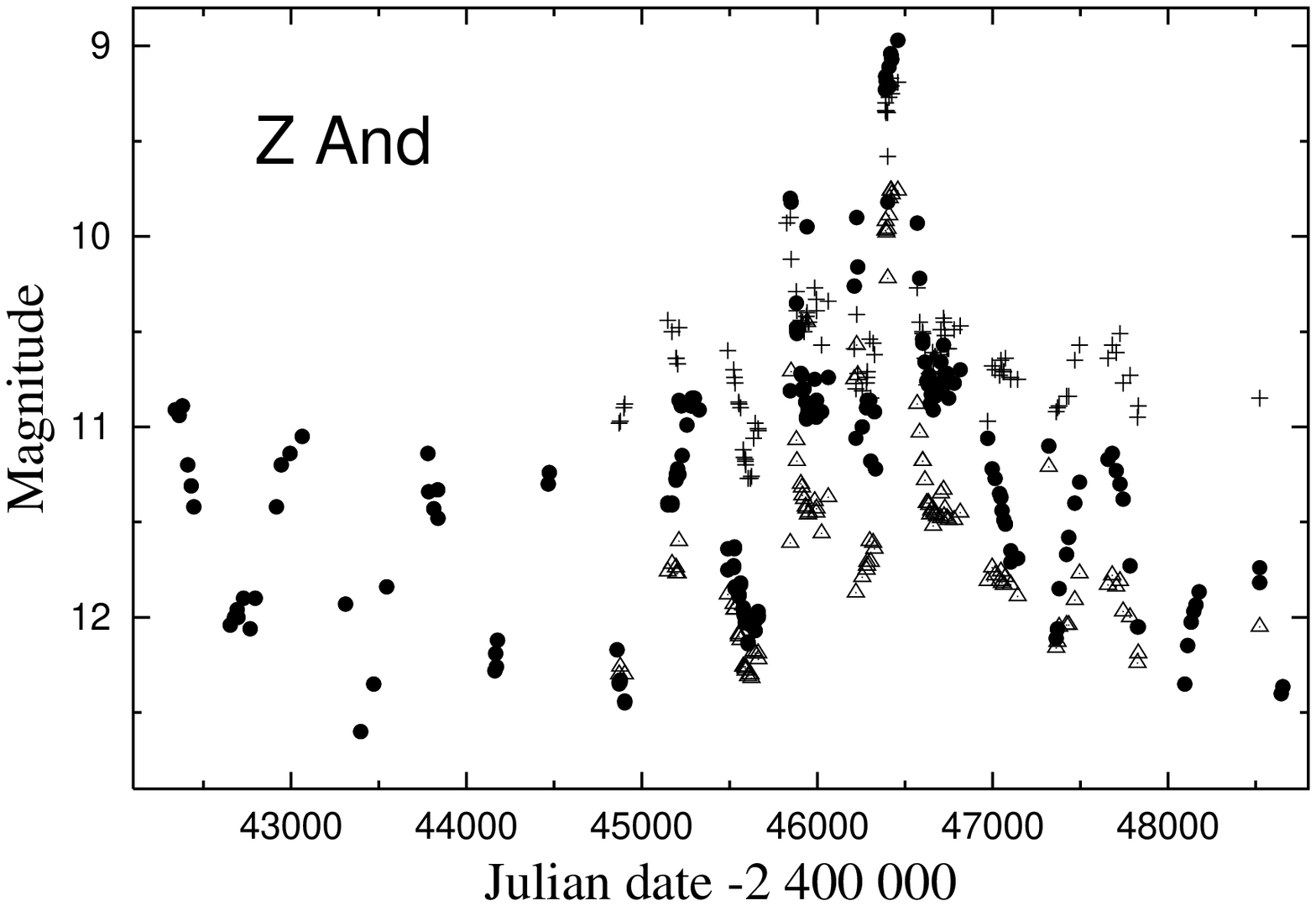,width=6cm,angle=-90}}}
\caption[]{
Examples of periodic wave-like variation in the light curves of 
selected symbiotic binaries. Symbols, $\bullet$, $\bigtriangleup$ 
and $+$, represent observations in the $U$, $B$ and $V$ band, 
respectively.
}
\end{figure*}

\subsection{On the nature of the wave-like variation}

\subsubsection{A reflection effect}

Originally, a reflection effect was suggested as responsible for 
the wave-like modulation in the light curve of AG\,Peg [3,5]. 
In this model the hot star irradiates
and heats up the facing giant's hemisphere that causes variation in
the star's brightness when viewing the binary at different orbital 
phases. This natural explanation has been adopted by many authors 
(e.g.: [23,34]), and recently approached 
quantitatively [39,40]. However, the observed 
large amplitude makes it difficult to get the model of the reflection 
effect consistent with observations. 

In the model of the reflection effect as approximated by [50], 
the upper limit (i.e. for $i = 90^{\circ}$) of the magnitude difference 
between the two hemispheres, $\Delta m_{\rm max}$, can be expressed as 
\begin{equation}
  \Delta m_{\rm max} = - 2.5\log (1 + 2 L_{\rm RE}/L_{\rm g}), 
\end{equation}
where $L_{\rm g}$ is the luminosity of the giant and $L_{\rm RE}$ 
represents a fraction of the hot star luminosity, $L_{\rm h}$, 
impacting the giant. For the separation of the stars, $A \gg R_{\rm g}$, 
$L_{\rm RE} = (R_{\rm g}/2A)^{2}\times L_{\rm h}$, and Eq. 1 reads 
as 
\begin{equation}
  \Delta m_{\rm max} = - 2.5\log (1 + \beta /2), 
\end{equation}
where the parameter 
\begin{equation}
  \beta=\frac{R_{\rm g}^{2}}{A^{2}}\frac{L_{\rm h}}{L_{\rm g}} 
      ~ =~ R_{\rm A}^{2} \frac{L_{\rm h}}{L_{\rm g}} 
\end{equation}
measures the strength of the illuminating radiation field relative to 
that of the giant. So the amplitude of light curves caused by 
the reflection effect is determined only by the parameter $\beta$. 
However, the parameter $\beta$ derived from observations can produce 
a maximum magnitude difference $\Delta m_{\rm max} < 0.1$\,mag, which 
is far below the observed quantities [50]. On the other hand, 
the reflection effect (Eq. 2) requires $\beta \ge 1 - 10$ to match 
the observed amplitudes. Even a more rigorous approach made 
by [39,40] does not provide a satisfactory explanation of 
the observed maximum differences in broadband magnitudes for 
realistic parameters $\beta \approx 0.1 - 0.01$. 

Another simplified approach to the reflection effect assumes a clear 
heating, i.e. the infalled light on the giant's hemisphere is absorbed 
and transformed into the heating up it. In this case the temperature 
of the irradiated hemisphere increases in maximum by a few $\times$ 10\,K 
for typical quantities of a symbiotic binary (see Eq. B.9 of [39]. 
Such small temperature difference between the two hemispheres cannot 
cause the observed amplitude in the light curves (e.g.: [3]). 
In a more realistic case, in which the giant's wind is ionized by 
the hot component radiation, only photons not capable of ionizing 
hydrogen ($\lambda > 912$\,\AA) can penetrate into the H\I\ zone at 
a vicinity of the giant's photosphere (Sect. 2.2.2, Fig. 3). 
In addition, a part of them (mainly those from the hot stellar source 
at the far-UV) are Rayleigh scattered by the neutral part of 
the red giant wind (particle density $n \sim 10^{10}\rm cm^{-3}$). 
As a result the real heating effect will probably be negligible. 
This is in agreement with observations of [44], who
did not find any irradiation effect in the red giant spectrum of SY\,Mus,
although its visual light curve varies with an amplitude of 0.6\,mag.

Finally, occultation of a bright gaseous region by the stellar disk
of the giant in the system was recently considered to explain the light 
variation in the $U$ band of AG\,Dra [59]. 
The nature of this emission region is seen in photoionization of 
the giant's wind. The authors assumed that this region is located 
most probably around the hemisphere of the giant facing the hot component. 
In the cases of a sinusoidal shape of the orbitally-related light variation 
(e.g. V1329\,Cyg, AG\,Peg, AG\,Dra, Z\,And; Fig. 2), about 50\% of the 
light, which is subject to variation, should still be occulted at the 
orbital phases 0.25 and 0.75, when viewing the binary from its side. 
However, the extra emission region is physically displaced from 
the giant's photosphere (the H\I/H\I\I\ boundary lies at 0.33\,$A$ from 
the giant in AG\,Dra on the line connecting the stars [50]), therefore 
a simple geometrical occultation cannot be responsible for such large 
loss of the light at these phases, mainly for the objects with 
$i\, \ll\,90^{\circ}$.  
%
\begin{figure}
  \centering
  \centerline{\hbox{
  \psfig{figure=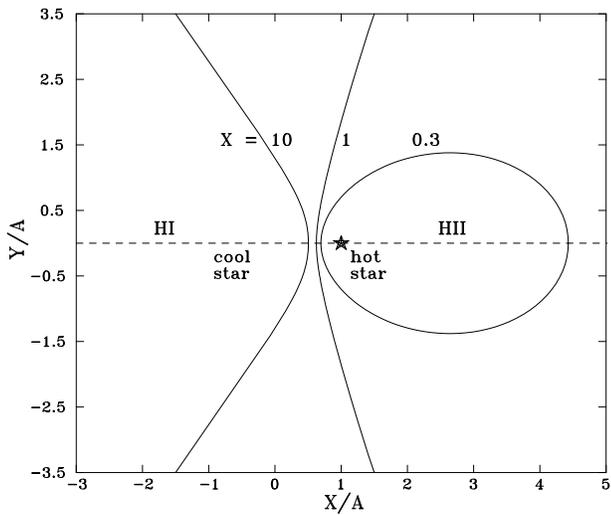,width=8cm}}}
  \caption{
   The H\I/H\I\I\ boundary calculated for 
   $X$ = 0.3, 1, 10, the stellar wind model characterized by the parameters 
   $\gamma$ = 2.5 and $R_{\rm g}/A$ = 0.28 (see [50] for more detail).
}
\end{figure}
\subsubsection{An ionization model and the wave-like variability}

All these reflection(-like) effects mentioned above assume the extra 
emission region to be localized at/on the giant's photosphere. This 
implies two main problems in explanation of the observed large 
magnitude variations: 
(i) A low emission of geometrically small region, and 
(ii) its localization under conditions of symbiotic binaries. 
Therefore we outline the nature of the orbitally-related changes in 
the optical/near-UV continuum within an ionization model of symbiotic 
binaries, i.e. within the basic model of symbiotic stars as introduced 
in section 1. \\
{\underline{\sf A zero-level ionization model.}} 
The extent of the ionized zone can be obtained from a parametric equation
\begin{equation}
   f(r,\vartheta) - X = 0,
\end{equation}
the solution of which defines the boundary between neutral and ionized
gas at the orbital plane determined by a system of polar coordinates, 
$r,\vartheta$, with the origin at the hot star [45]. By other words 
the ionization boundary is defined by the
locus of points at which ionizing photons are completely consumed along
paths outward from the ionizing star. The function $f(r,\vartheta)$ was 
treated for the first time by [45] for a steady state 
situation (a constant velocity of the wind and no binary rotation) and 
pure hydrogen gas. The parameter $X$ is given mainly by the binary 
properties -- separation of the components, number of hydrogen 
ionizing photons, terminal velocity of the wind and the mass-loss 
rate (for details see [36,45]).
The particle density of the ionized material is given by the velocity
distribution of the giant's wind, which is here assumed to be of 
the form 
\begin{equation}
 v_{\rm wind} = v_{\infty}(1- R/r)^{\gamma}, 
\end{equation}
where $r$ is the distance from the center of the cool star, $R$ is 
the origin of the stellar wind (\,$\approx $\,the radius of the giant) 
and $v_{\infty}$ is the terminal velocity of the wind. 
The parameter $\gamma$ characterizes an acceleration of the wind. 
A larger $\gamma$ corresponds to a slower transition to $v_{\infty}$.
Figure 3 shows examples of the H\I/H\I\I\ boundary for 
$X$ = 0.3, 1, 10 and the parameter $\gamma$ = 2.5 in the wind model. \\
\begin{table}[t]
\begin{center}
\caption{Observed and calculated $EM$ for selected objects.}

\vspace{1mm}

\begin{tabular}{cccc}
\hline
Object & $F_{\lambda}^{\rm obs}/10^{-13}$ & $EM_{\rm obs}$ & $EM_{\rm M}$ \\
       &                                  &  [cm$^{-3}$]   & [cm$^{-3}$]  \\
\hline
BF\,Cyg & 5.0$^{a}$       & 3.8$\times 10^{60}$       & 2.9$\times 10^{60}$  \\
BF\,Cyg & 2.7 - 7.9$^{b}$ & 3.6 - 10$\times 10^{60}$  &                      \\
AG\,Dra & 2.4 $^{a}$      & 1.5 $\times 10^{59}$      & 1.5$\times 10^{59}$  \\
AG\,Dra & 0.9 - 2.2$^{b}$ & 0.5 - 1.3$\times 10^{59}$ &                      \\
AX\,Per & 2.0$^{a}$       & 2.1$\times 10^{59}$       & 1.2$\times 10^{59}$  \\
AX\,Per & 1.0 - 2.5$^{b}$ & 1.8 - 4.6$\times 10^{59}$ &                      \\
V443\,Her & 2.1$^{a}$     & 3.0$\times 10^{59}$       & 4.2$\times 10^{59}$  \\
V443\,Her & 1.1 - 2.1$^{b}$ & 2.7 - 5.2$\times 10^{59}$ &                    \\
\hline
\end{tabular}
\end{center}
$a$ - from the energy distribution in the spectrum at 
      $\lambda$3646$^{-}$\AA. \\
$b$ - from the dereddened $U$-magnitude at minimum and maximum,
      respectively. \\
Flux in units of erg\,cm$^{-2}$\,s$^{-1}$\,\rm \AA$^{-1}$
\end{table}
%
%
\begin{figure}
  \centering
  \centerline{\hbox{
  \psfig{figure=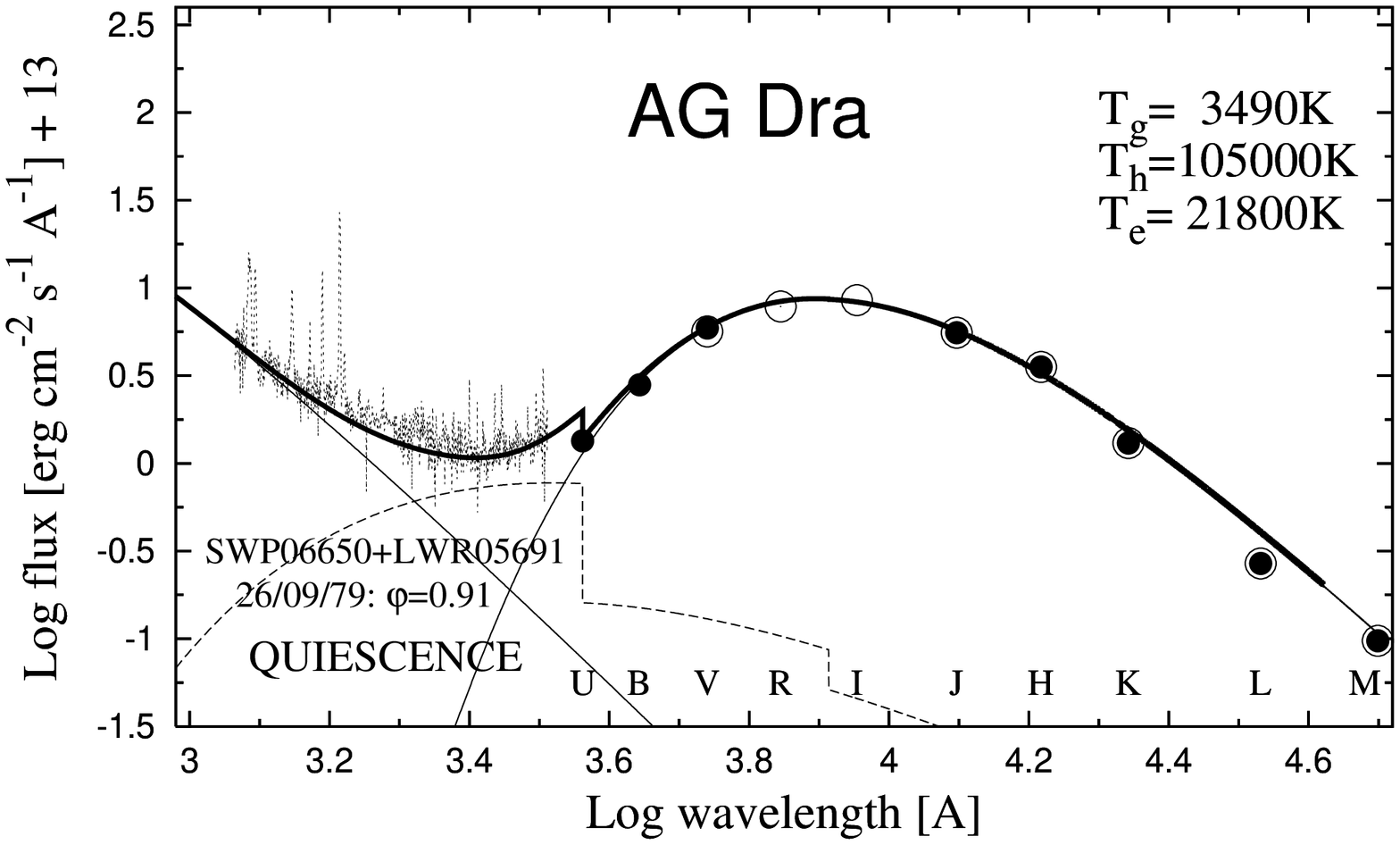,width=8cm,angle=-90}}}
\vspace*{0.2cm}
  \centerline{\hbox{
  \psfig{figure=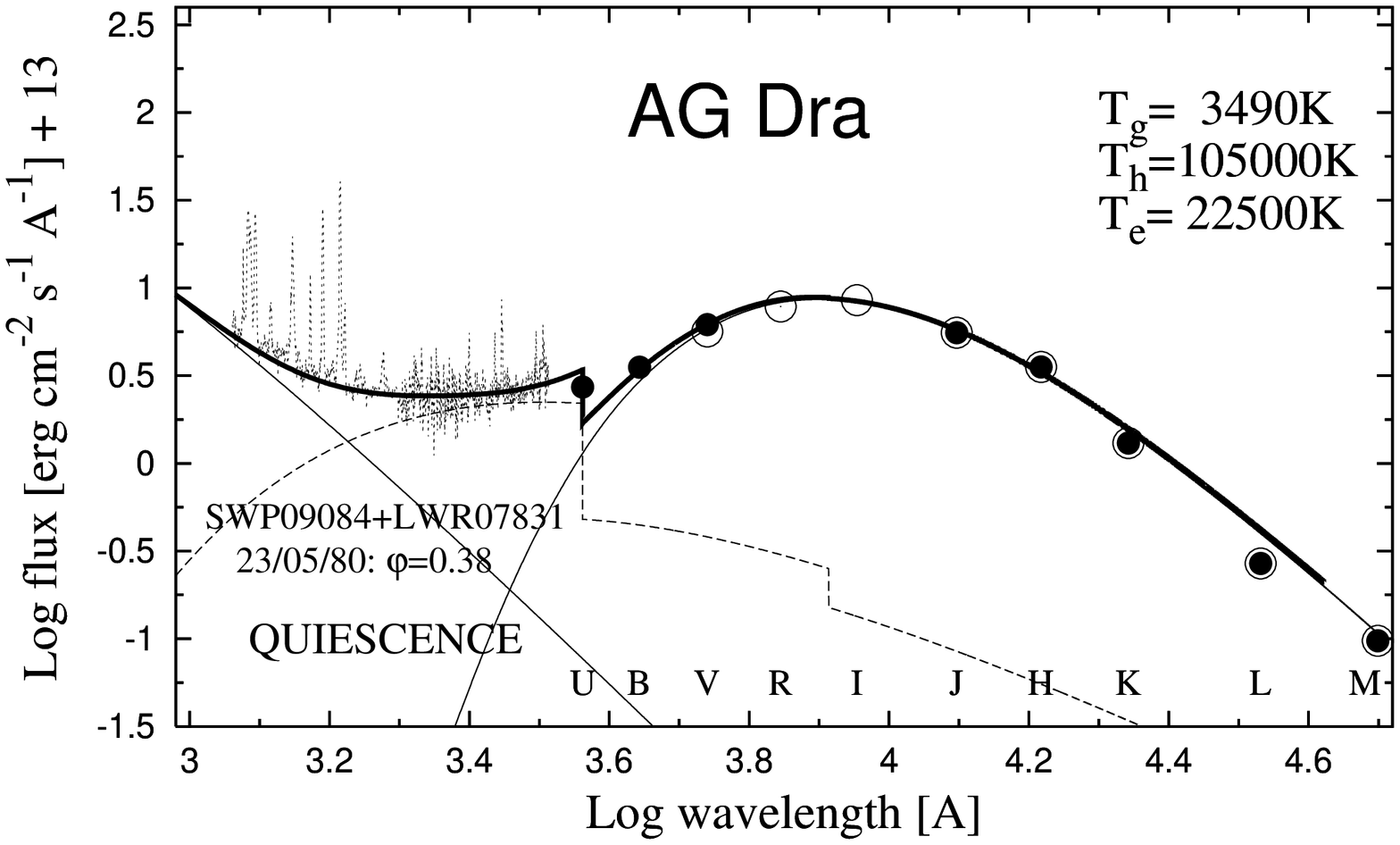,width=8cm,angle=-90}}}
\caption{
Example of the SED of AG\,Dra at the orbital phase 0.91 (top) and 
0.46 (bottom), which demonstrates the orbital variability in 
the nebular emission. 
}
\end{figure}
\noindent
{\underline{\sf Amount of emission from the model and observations.}}\\
Here we show that the nebular emission in the continuum produced by 
this model can balance that given by observations. 
The nebular flux largely depends on the number of hydrogen 
recombinations, and is proportional to $\int \!n_{+}n_{\rm e}\,dV$, 
(the so called emission measure - $EM$); $n_{+}$ and $n_{\rm e}$ is 
the concentration of ions (protons) and electrons, respectively. 

(i) {\em Observations:} The quantity of the $EM$ can be estimated, 
for example, from the measured flux, $F^{\rm obs}_{\lambda}$ 
($\rm erg\,cm^{-2}\,s^{-1}\,\AA^{-1}$), of the nebular continuum 
at the wavelength $\lambda$, as
\begin{equation}
  EM_{\rm obs} = 4\pi d^{2} F^{\rm obs}_{\lambda}/ \varepsilon_{\lambda},
\end{equation}
in which $d$ is the distance to the object, $\varepsilon_{\lambda}$ is
the volume emission coefficient. The nebular flux can be obtained from 
the energy distribution in the spectrum. Its upper limit can also be 
estimated from the dereddened $U$-magnitude if the nebular continuum 
dominates the optical. 

(ii) {\em Model:} The source of the nebular radiation in the model 
is the ionized region, in which the rate of ionization/recombination 
processes is balanced by the rate of photons, $L_{\rm ph}$ 
(photons\,s$^{-1}$), capable of ionizing the element under 
consideration. In the case of pure hydrogen we can write 
the equilibrium condition as 
$L_{\rm ph} = \alpha_{\rm B} \!\int_{V} \!\!n_{+}(r)n_{\rm e}(r)\,dV,$ 
where $\alpha_{\rm B}$ ($\rm cm^{3}\,s^{-1}$) is the total hydrogenic 
recombination coefficient and $L_{\rm ph}$ is given by the luminosity 
and temperature of the hot star. So the quantity of the emission 
measure given by the model ($EM_{\rm mod}$) can be written as
\begin{equation}
 EM_{\rm mod} =  L_{\rm ph}/ \alpha_{\rm B}.
\end{equation}
Having independently determined $L_{\rm ph}$ we compare the $EM$ 
gi\-ven by observations (Eq. 6) and that required by the ionization 
model (Eq. 7). Results for some objects are in Table 1, which 
demonstrates that 
\begin{equation}
        EM_{\rm obs}~\sim ~EM_{\rm mod}. 
\end{equation}
So the amount of emission produced by the ionization model is consistent
with observations. \\
%
\begin{figure}[t]
  \centering
  \centerline{\hbox{
  \psfig{figure=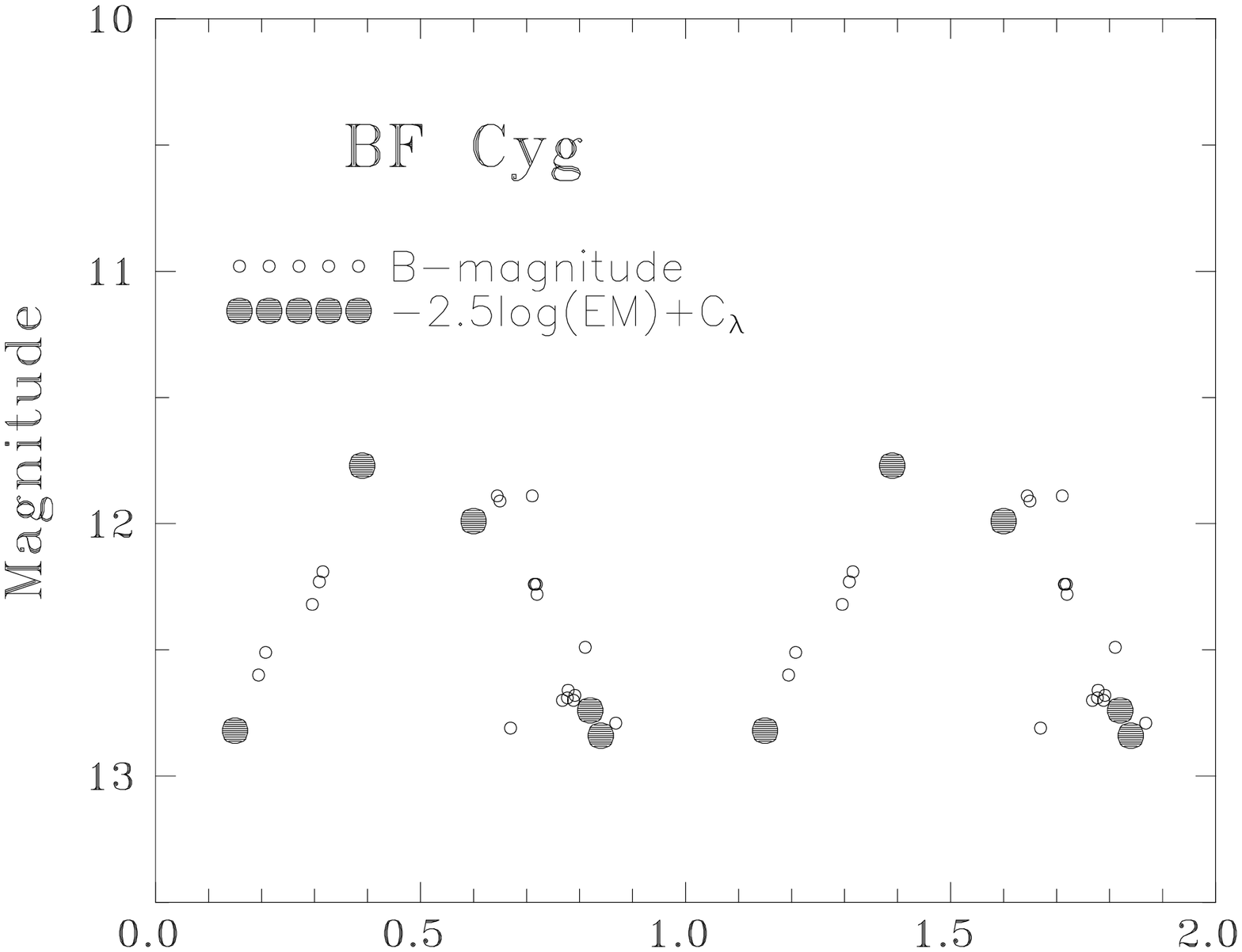,width=8cm}}}
  \centerline{\hbox{
  \psfig{figure=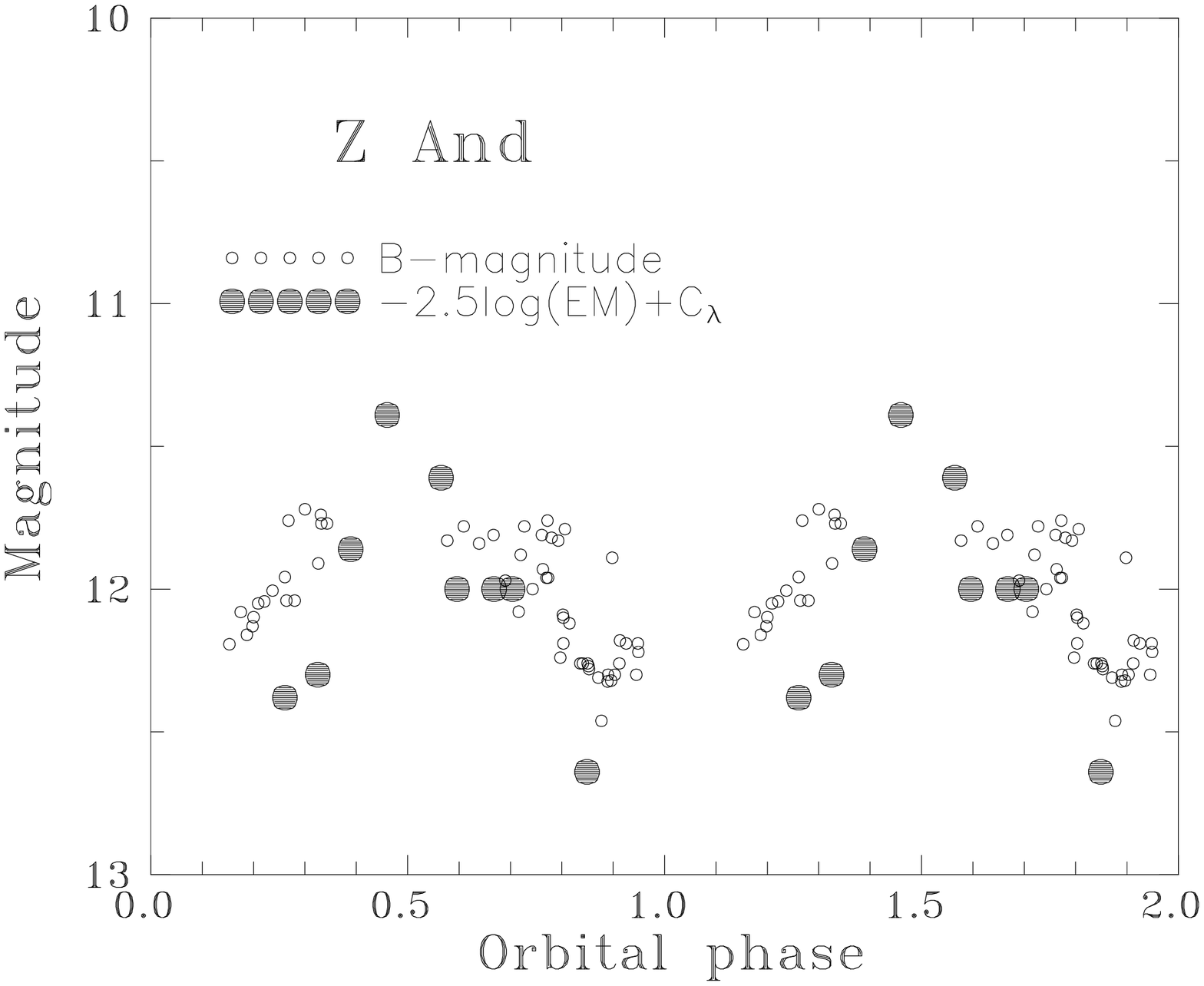,width=8cm}}}
\caption{
Top: Variation in the $EM$ of BF\,Cyg as a function
of the orbital phase. The data [31] were converted
into the $B$-magnitudes according to Eq. 9. Compared is the light curve 
in the $B$ band obtained photometrically during the same period [21]. 
Bottom: The same as the top, but for Z\,And. Measurements of the $EM$ 
were taken from [11], and photometric $B$-magnitudes from Fig. 1, 
but omitting the active phase. These results show that
the variation in the $EM$ is fully responsible for that in 
the light curves (from [50]). 
}
\end{figure}
%
%
{\underline{\sf Variation in the $EM$ and light curves.}}
Here we show that the variation in the $EM$ is responsible for 
the investigated wave-like variation in the light curves. 

The quantity of the $EM$ also varies as a function of the orbital phase 
(e.g.: [11,31]). 
In Fig. 4 we show an example of the SED of AG\,Dra at the orbital 
phase 0.46 and 0.91, i.e. close to its maximum and minimum, respectively. 
This demonstrates that the nebular continuum emission 
is subject to variation with the orbit of the binary. 
To compare the observed variability in $EM$ to that in the light curves, 
we express Eq. 6 in the scale of magnitudes 
($m_{\lambda}=-2.5\log(F_{\lambda})+q_{\lambda}$) as 
\begin{equation}
 m_{\lambda} = -2.5\log(EM) + C_{\lambda}, 
\end{equation}
where 
\begin{equation}
 C_{\lambda} = q_{\lambda} - 2.5\log\big(\frac{\varepsilon_{\lambda}}
               {4\pi d^{2}}\big),
\end{equation}
in which the constant $q_{\lambda}$ defines magnitude zero. 
Using Eq. 9 we constructed the light curves from the measured values of 
the $EM$ in BF\,Cyg and Z\,And and compared them to those obtained 
photometrically. We adopted distances $d$ = 4.6 and 1.12\,kpc for BF\,Cyg 
and Z\,And, respectively. Figure 5 shows that the variation in 
the $EM$ follows well that observed in the light curves (see [50] for 
details). Finally, we need to answer the question: "What is the origin 
of the periodic wave-like variability?" So, \\
%
{\underline{\sf Why does the emission measure vary?}}
To explain the orbi\-tally-related variation, the nebula has to be 
partially optically thick and of a non-symmetric shape (Sect. 3.2.1) 
to produce different contributions of its total emission into 
the line of sight at different orbital phases. 
In our simple ionization model the opacity, $\kappa$, of the ionized 
emission medium decreases with the distance from the cool star, 
since $\kappa \propto n \propto r^{-2}$ (i.e. the parts nearest to 
the giant's photosphere will be most opaque).
From this point of view, shaping of the optically thick portion of 
the H\I\I\ zone will be given by the parameter $X$ in our ionization 
model (Eq. 4, Fig. 3). 
To characterize the basic shape of the {\em observed} light curves 
[47] introduced a parameter $a$ as
\begin{equation}
 a = \frac{m(0) - m(0.25)}{\Delta m_{\rm max}},
\end{equation}
where $m(0)$ and $m(0.25)$ are the magnitudes at the orbital phases 
0 and 0.25, respectively, and $\Delta m_{\rm max}$ is the amplitude of 
the light curve. The shape of the light curve resembles a sinusoid 
for $a = 0.5$, but $a > 0.5$ implies a broader maximum than minimum.
[50] found a relationship between the parameter $a$ and 
the parameter $X$ (Fig. 6). This supports a connection between 
the shape of light curves and the extent of symbiotic nebulae. 
Accordingly, [50] suggested a qualitative description 
on how the observed profile of light curves could be produced 
within the ionization model mentioned above:

(i) In the extensive emission zone ($X \ge 10$), the partially 
optically thick portion of the H\I\I\ region has the geometry of 
a cap on the H\I/H\I\I\ boundary around the binary axis. This 
mimics the reflection effect, but the emission region causing the 
light variation is physically displaced from the giant's surface. 
In this case the light curve profile is a sinusoid characterized 
by the parameter $a \le 0.5$ (e.g. V1329\,Cyg, Z\,And, AG\,Dra). 

(ii) In the case of an oval shape of the H\I\I\ zone (Fig. 3, 
a small parameter $X$), its total emission will be attenuated more 
at positions of the inferior and superior conjunction of the cool
star (the orbital phase $\varphi$ = 0 and 0.5, respectively) than 
at positions of $\varphi$ = 0.25 and 0.75, respectively. Such  
apparent variation in the $EM$ can produce a primary as well as 
secondary minimum in the light curve. This type of the light
curve profile corresponds to the parameter $a \sim 1$ 
(e.g. EG\,And).
A gradual opening of the H\I\I\ zone (approximately      
$0.3 < X < 1$) will make it optically thinner behind the hot     
star (outside the binary around $\varphi$ = 0.5). Therefore 
the secondary minimum will become less pronounced or flat, and/or 
a maximum at $\varphi \sim 0.5$ can arise. The light curve profile 
here should be characterized by the parameter $a > 0.5$ (e.g. CI\,Cyg, 
AX\,Per). 
However a quantitative approach to determine opacities of a more 
realistic structure of the ionized region in symbiotic binaries 
is required. 
%
%
\begin{figure}
  \centering  
  \centerline{\hbox{
  \psfig{figure=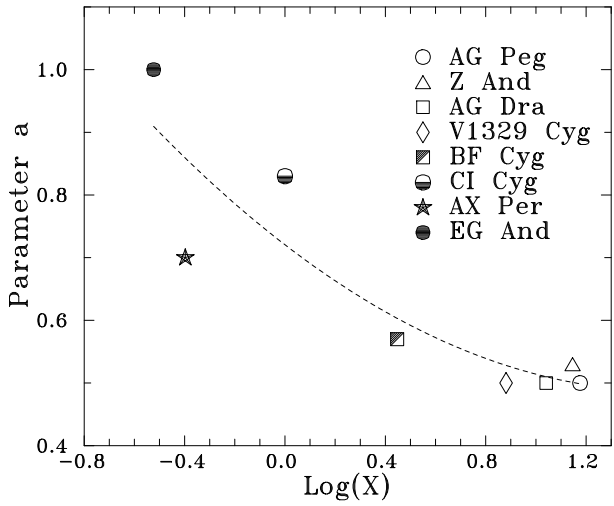,width=8cm}}}
\caption{
A correlation between the shape of the light curve characterized
by the parameter $a$ (see the text) and the extent of the symbiotic   
nebula given by the parameter $X$ (from [50]). 
}
\end{figure}

\section{Ionization during transition periods}

In this section we discuss the role of ionization in symbiotic 
binaries during their transition from active to quiescent phases. 
Based on analysis of the long-term photometric observations 
of a group of classical symbiotic stars, it was shown that a transition 
between different levels of activity produces a systematic variation in 
the $O-C$ residuals and thus an apparent change in 
the orbital period [48]. 

\subsection{Systematic variation in the O-C residuals}

Here we demonstrate this effect on historical light curves of 
classical symbiotic stars BF\,Cyg and AG\, Peg. First, we determine 
positions of the observed minima in their light curves. Second, we 
construct the $O-C$ diagram using a reference (spectroscopic) 
ephemeris. By this way we get a relative position between the time 
of the inferior conjunction of the cool star and the observed light 
minimum, i.e. the location of the main emission source in the binary 
with respect to the line connecting the stars. 

{\bf BF\,Cyg} ($P_{\rm orb}$=757.3 days): 
Top left panel of Figure 7 shows the historical light curve with 
the $O-C$ diagram of BF\,Cyg. As the reference ephemeris we adopted that 
given by all the primary minima measured by [48], 
\begin{equation}
JD_{\rm Min} = 2\,411\,268.6 + 757.3(\pm 0.6)\times E,
\end{equation}
which is identical (within uncertainties) with the spectroscopic 
e\-phe\-meris of [10]. 
A systematic variation in the $O-C$ residuals is clearly seen. This 
behaviour was already noted by [22]. An increase before 
the 1920 bright stage (E\,=\,1 to 11) corresponds to a period of 770 days, 
while the subsequent decrease (E\,=\,12 to 24) indicates a shorter period 
of 747.5 days. 
The same type of variability appeared again during the recent, 1989 active
phase. Observed changes in both the position and the shape of the minima  
are illustrated in the top right panel of Fig. 7. During the transition 
{\em from the active phase to quiescence} (the $A\rightarrow Q$ transition), 
a systematic change in the minima positions at E = 49 to 51 corresponds to 
a period of 730 days. The following minima (E = 52, 53 54), which were 
observed during quiescent phase, do not exhibit any systematic change 
in their position. 
During the transition {\em from the quiescent to the active phase} 
($Q\rightarrow A$), a significant change in the $O-C$ values by jump
of +130 days was observed. The minima at E = 45 to E = 49 indicate 
an apparent period of 794 days. 
%
%
\begin{figure*}[t]
  \centering
  \centerline{\hbox{
  \psfig{figure=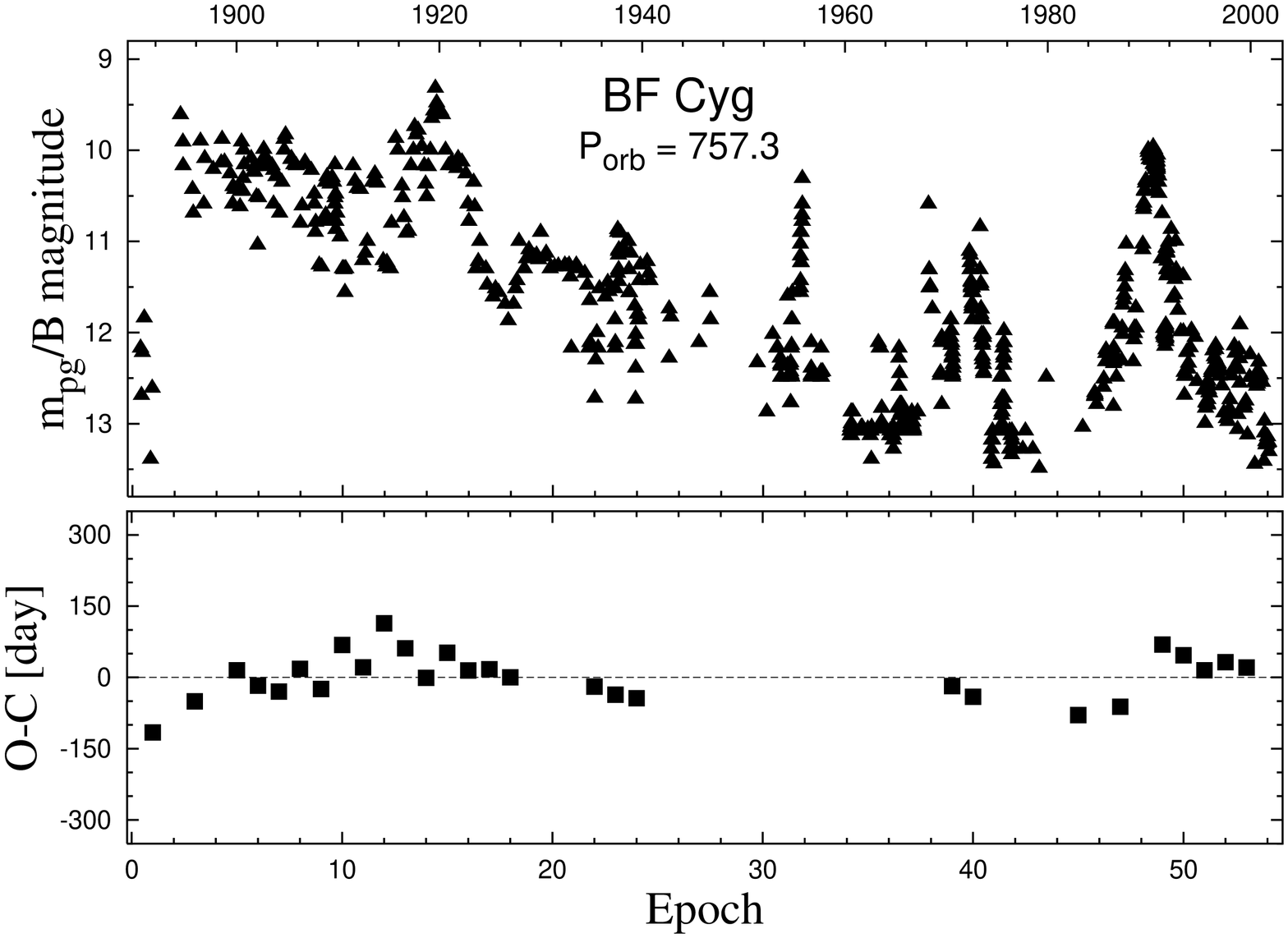,width=8cm,angle=-90}
\hspace*{0.1cm}
  \psfig{figure=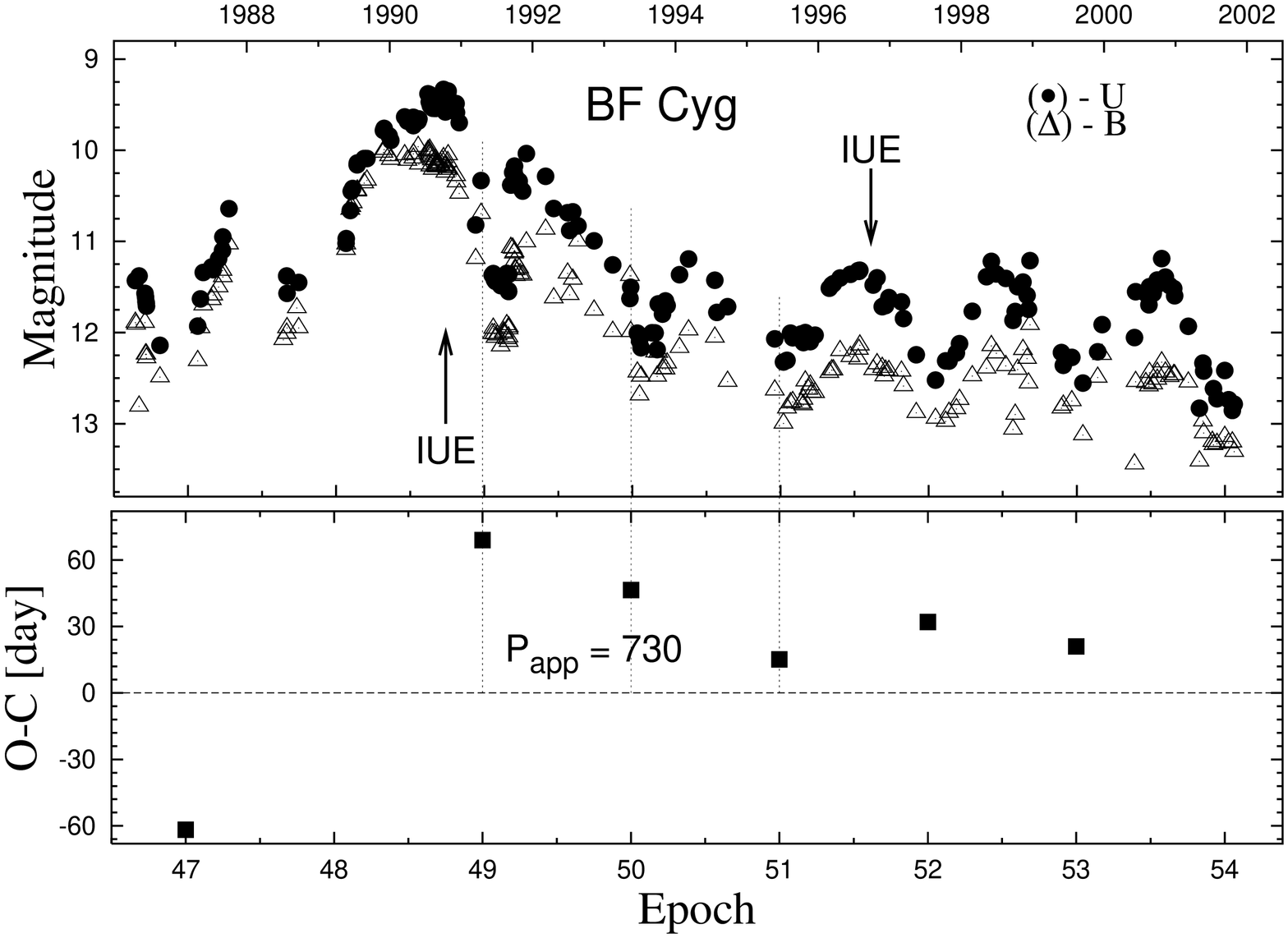,width=8cm,angle=-90}}}
\vspace*{0.1cm}
  \centerline{\hbox{
  \psfig{figure=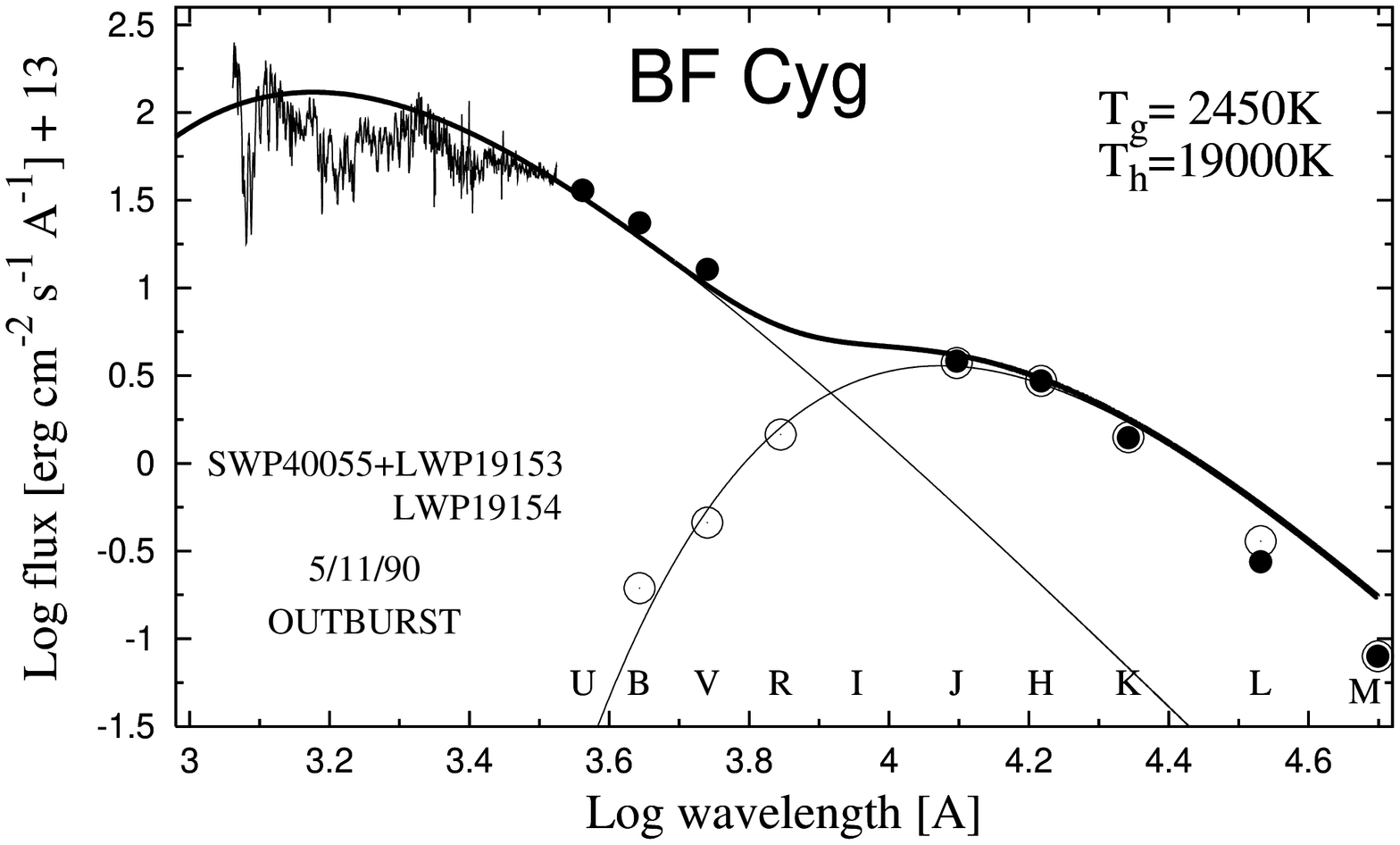,width=8cm,angle=-90}
\hspace*{0.1cm}
  \psfig{figure=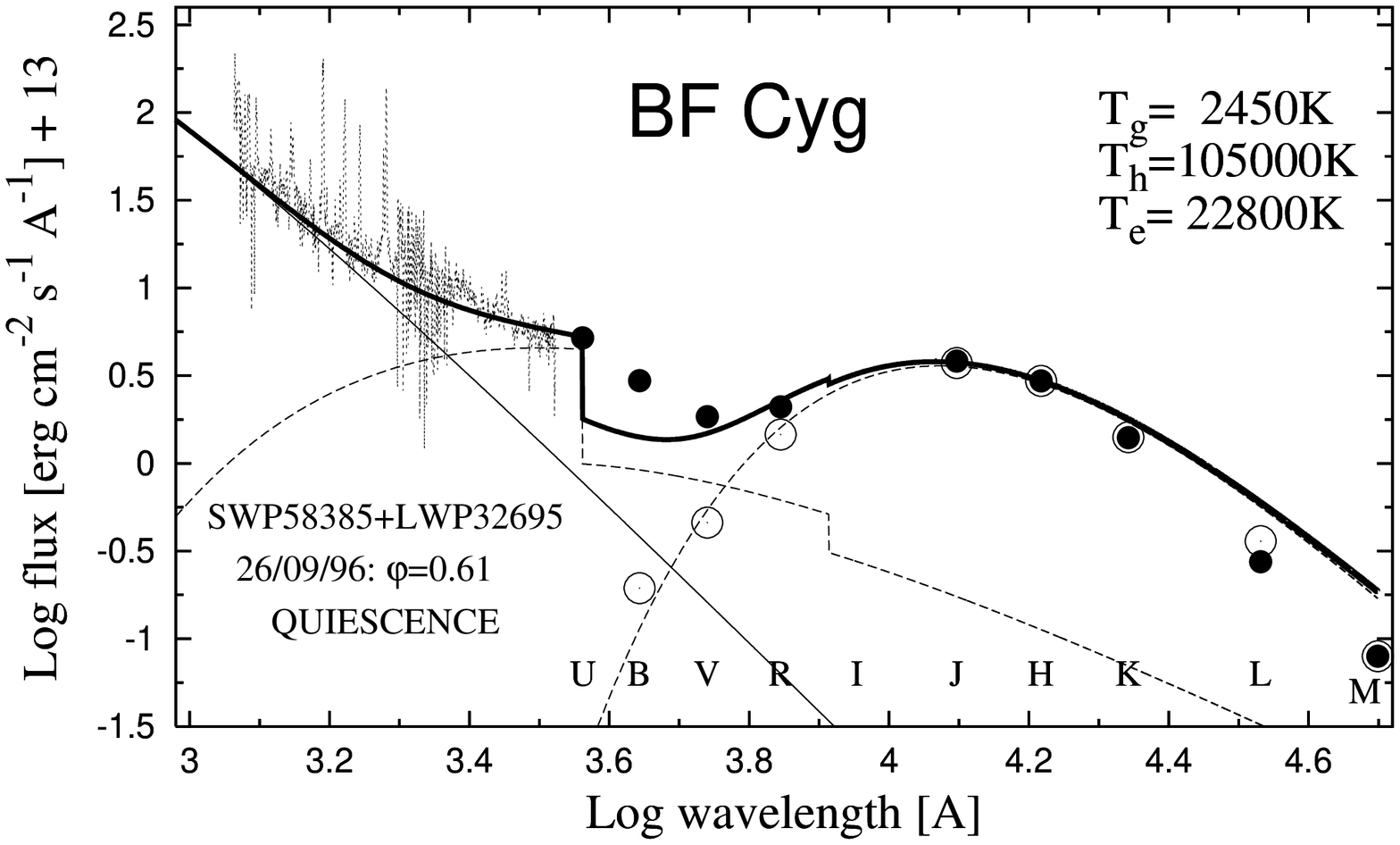,width=8cm,angle=-90}}}
\caption{
Top: Historical and recent light curves of BF\,Cyg with the $O-C$ 
diagrams for the observed minima. During the recent transition to 
quiescence (E = 49, 50, 51), an apparent period of 730 days was 
indicated. 
Bottom: The SED during the active and the quiescent phase. This 
together with the change in the shape and position of the minima 
demonstrate the change in the geometry and location of the symbiotic 
nebula during the $A\rightarrow Q$ transition. 
}
\end{figure*}

{\bf AG\,Peg} ($P_{\rm orb}$=812.6 days):
As the reference ephemeris for timing of the inferior conjunction 
of the cool giant in AG\,Peg we can adopt that of [48] 
\begin{equation}
JD_{\rm sp. conj.} = 2\,427\,664.2(\pm 10) + 812.6(\pm 1.8)\times E,
\end{equation}
which is based on about 20 orbital cycles during 1945 -- 1992. 

From $\sim$1940 the light curve developed a periodic wave-like
variation [28]. The periodic variability has been very
intensively studied. The real shifts of the observed minima from 
the predicted positions were often noted [4,26,27]. As a result, 
many different periods ranging from $\sim$730 to $\sim$830 days 
were suggested ([25] and references therein). Generally, the older 
data gave a longer period (e.g.: [28]) than the more recent 
observations (e.g.: [12]).
Figure 8 shows the photographic and visual light curves from 1935.
Characteristic points (mostly maxima and minima) were taken from
photographic measurements [28], while the visual data
represent smoothed AFOEV (Association Francaise des Observateurs 
d'Etoiles Variables) estimates available from CDS. Positions of 
the observed minima define the ephemeris 
\begin{equation}
  JD_{\rm Min} = 2\,427\,495.9 + 820.3(\pm 0.8)\times E.
\end{equation}
To confirm the real difference between both the photometric and 
the spectroscopic period, we divided the available data set of 
radial velocities into two parts: 
(i) The old data from 1945.8 to 1973.8 (24 measurements), and 
(ii) the more recent data from 1978.5 to 1992.0 (58 measurements). 
Then we solved circular orbits for each data set separately 
with fixed period of 820.3 days. Both solutions differ from each 
other only in the time of spectroscopic conjunction corresponding 
to an average shift of 0.15\,$P_{\rm orb}$ between them (Fig. 9 top). 
However, the phase diagram of all radial velocities constructed for 
the elements in Eq. 13 does not display any systematic shift 
(Fig. 9 bottom). This means that the {\em photometric} period
is inconsistent with the {\em orbital} period, and thus represents 
an apparent period in the system. The $O-C$ residuals display 
a systematic increase along a gradual decrease of the star's brightness. 
Such behaviour reflects a longer apparent period than the orbital one. 

\subsubsection{Common properties}

By this way [48] analyzed other 5 symbiotic systems 
(CI\,Cyg, V1329\,Cyg, Z\,And, AG\,Dra and YY\,Her) and found 
following two main characteristics in the variation of 
the $O-C$ residuals: 

1. {\em Systematic variation in positions of the minima is connected
with a variation in the energy distribution of the hot component
radiation.} 

The effect is very striking when also the nature of the hot continuum
changes -- from a blackbody during outburst to a nebular radiation 
in a quiescence. In this case the apparent period is shorter than 
the orbital one. Figure 7 demonstrates this case for the eclipsing 
system BF\,Cyg on its SED during the 1990 outburst and the following 
quiescent phase. 

The case, in which no change in the nature of the hot continuum 
was observed, we demonstrated on the example of AG\,Peg (Fig. 8). 
At the beginning of its nebular phase, when the wave-like variation 
developed in the light curve (from $\sim$\,1940), the minima were 
shifted by about -100 to -200 days from their prediction 
by Eq. 13. Then, the following minima were appearing systematically 
closer to the time of conjunction. In such the case the apparent period 
is larger than the orbital one.
Also here behaviour in the $O-C$ residuals was followed by 
a gradual decline of the star's brightness. 

2. {\em Separation between the minima correlates with the velocity of
the brightness variation.}

Generally, a fast and large change in the period is indicated during
$Q \rightarrow A$ transitions, when the brightness changes rapidly.
Contrary, $A\rightarrow Q$ transitions, during which the star's brightness
declines slowly, produce a smaller change in the period. 
[48] determined a relation between parameters 
$\Delta m /\Delta t$ and $\Delta P /P_{\rm orb}$ characterizing 
the velocity of the change in the star's brightness and 
the corresponding phase shift in the period, respectively, for 
different transition epochs in a larger sample of symbiotic stars as 
\begin{equation}
\frac{\Delta P}{P_{\rm orb}} \, \sim \, 134 \frac{\Delta m}{\Delta t},
\end{equation}
where the parameter $\Delta m /\Delta t$ is in $mag/day$. 
%
%
%
\begin{figure}
\centering
   \centerline{\hbox{
   \psfig{figure=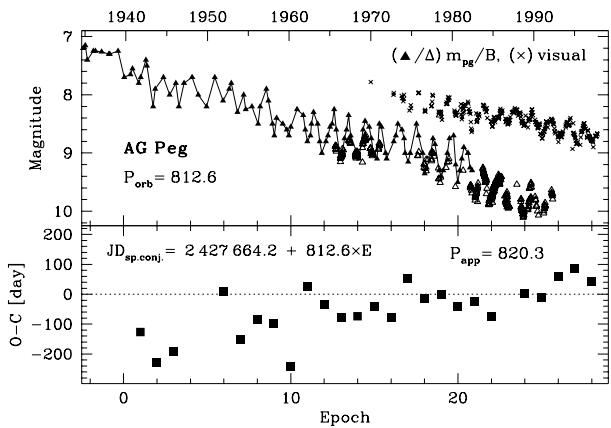,width=8cm}}}
   \caption{
Top: Compiled photographic/B and visual light curves of AG\,Peg 
as recorded from 1935. 
Bottom: The $O-C$ diagram for the observed minima. A gradual 
increase in the $O-C$ values indicates an apparent period of 
820.3 days. 
}
\end{figure}
%
%
\begin{figure}[t]
\centering
\centerline{\hbox{
   \psfig{figure=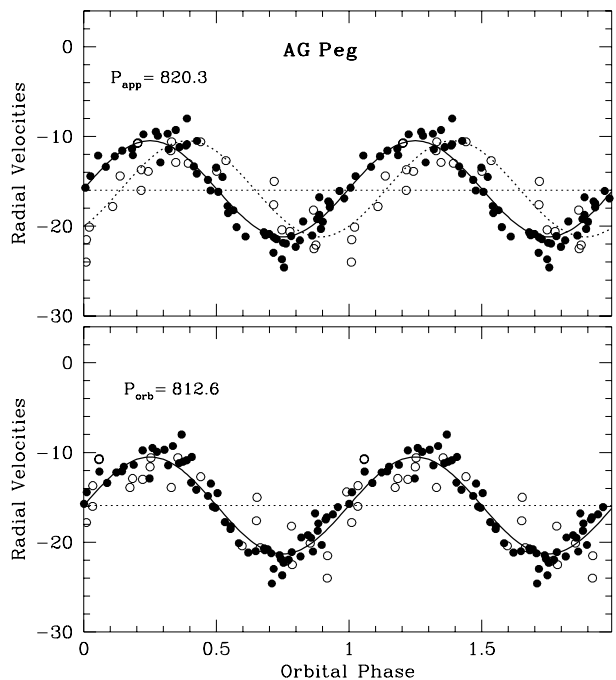,width=8cm}}}
   \caption{
Top: The old, 1945.8 - 1973.8, radial velocities (open circles) 
and the more recent, 1978.8 - 1992.0, measurements (full circles) 
are shifted by $\sim$0.15\,$P_{\rm orb}$ relative to each other 
when folded with the photometric period of 820.3 days. 
Bottom: No systematic shift can be seen when the data are folded 
with the orbital period of 812.6 days.
}
\end{figure}

\subsection{Principle of apparent orbital changes}

Generally, the shape and position of a minimum in the light curve 
reflect the geometry and location of the main source of the optical 
continuum in the binary. 
In symbiotic binaries the nebular radiation often dominates 
the optical continuum during quiescence (cf. Fig. 1), but significantly 
changes during active phases (Fig. 7). Therefore the observed 
photometric variations are caused mainly by changes of the 
ionization structure in symbiotic binary due to outbursts, i.e. due 
to variation in the luminosity of ionizing photons. 
Basically, there can be recognized two different sources of 
the hot continuum in the symbiotic system -- a relatively cool 
and small pseudophotosphere around the hot component and the extended 
asymmetrical H\I\I\ zone -- dominating the optical spectrum during active
and quiescent phases, respectively (Fig. 7). Thus, during outbursts 
we can observe deep narrow minima -- eclipses -- at the position of 
the spectroscopic conjunction, while during quiescence the minima 
are broad, complex in profile, and shifted from the position
of the spectroscopic conjunction (here BF\,Cyg). 
A systematic change in timing of the minima can occur also if only 
the luminosity of ionizing photons changes, without any creation of 
a cool pseudophotosphere around the hot star, because this also produces
a systematic variation in the shape of the H\I\I\ zone (here AG\,Peg). 
However, the variation is not so drastic as can be observed in 
the former case.

\subsubsection{Asymmetric shape of the H\,II zone}

Observation of the systematic variations in the $O-C$ residuals 
suggests an asymmetrical shape of the H\I\I\ zone with respect to 
the binary axis to describe the observed apparent changes in 
the orbital period. There are also some observational arguments 
and calculations suggesting the asymmetrical shape of 
the H\I\I\ zone: 

(i) Direct observational evidence of an asymmetrical H\I\ (and thus
also H\I\I) region is represented by a wide 'eclipse', lasting from 
the orbital phase $\sim$0.9 to $\sim$1.6, in the far-UV region
caused by Rayleigh scattering in BF\,Cyg [38]. 

(ii) Asymmetric shape of the ionization front is suggested by
spectropolarimetric studies of [41,42]. The geometry of 
the H\I\I\ zone has no symmetry with 
respect to the binary axis to explain the observed polarization 
properties. In these models, due to the orbital motion, the ionization 
front in the orbital plane is twisted, going from the side
of the hot star that precedes its orbital motion, through the line joining
the components, to the front of the cool star against its motion. 

(iii) Asymmetrical nebular geometry is also suggested by hydrodynamical
calculations of the structure of stellar winds in symbiotic stars that
include effects of the orbital motion. The recent models [14,66,67]
are characterized 
by the S-shaped wind-collision zone, placed in the binary likewise 
as the ionization front described in the point (ii). 

(iv) Finally, [50] introduced a modification of the simplest 
steady state ionization model as introduced in Sect. 2.2.2 by including 
the orbital motion and a velocity structure of the giant's wind. 
In this model the H\I\I\ zone for small values of the $X$ parameter 
is also asymmetrically placed in the binary with respect to the 
binary axis. Figure 10 shows this case. 

Therefore we can assume that the shape of the optically thick 
fraction of the H\I\I\ region is prolonged in a direction between 
the stars and its projection into the orbital plane is inclined 
relatively to the binary axis, probably shaped as described in 
the point (ii). 
%
%
\begin{figure}[t]
\centering
   \centerline{\hbox{
   \psfig{figure=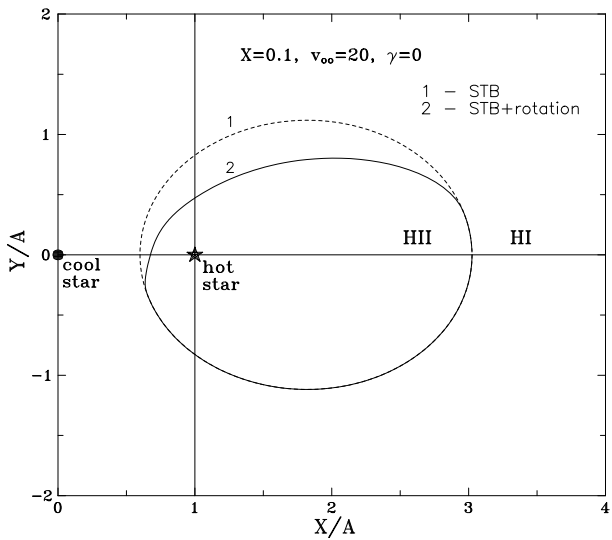,width=8cm}}}
  \caption{
The H\I/H\I\I\ boundary given by the steady state ionization model
(dashed line) and its modification due to the orbital motion
(solid line). This model was calculated for $P_{\rm orb}$ = 757 days, 
separation between the binary components $A$ = 465\,R$_{\odot}$ and 
the mass ratio $q = 6$. The binary rotates anti-clockwise 
([50]). 
}
\end{figure}

\subsubsection{The cases of BF\,Cyg and AG\,Peg}

A {\em gradual} decrease in the star's brightness results in
a {\em systematic} variation of the structure of the H\I\I\ region, 
which then produces the observed variation in the $O-C$ residuals 
and thus an apparent change in the period. Here we distinguish 
two situations: 

(i) A {\em decrease} of the total luminosity, as during nebular stage 
of AG\,Peg (the nature of the optical continuum is not changed), 
produces also a decrease of the luminosity of the hydrogen ionizing 
photons. This leads to a gradual {\underline{shrinking of the H\I\I\ zone}}, 
and thus the minimum can be expected to occur prior, but closer to 
the position of the spectroscopic conjunction than during the previous 
cycle. As a result a {\underline{larger apparent period}} than 
the orbital one is observed. 

(ii) An approximately {\em constant} hot star bolometric lumino\-si\-ty 
during a gradual decrease of the star's brightness throughout
the $A\rightarrow Q$ transitions of BF\,Cyg, produces an increase 
in the luminosity of the ionizing photons as the temperature of 
the hot star increases. This is connected with a significant change 
in the nature of the optical continuum  -- from a blackbody to 
a nebular radiation (Fig. 7), which causes an 
{\underline{expansion of the H\I\I\ zone}} that fits 
gradually its prolonged shape. The light minima, contrary to 
the case (i), will occur prior to, but systematically further 
from the time of spectroscopic conjunction. As a result, 
a {\underline{shorter apparent period}} than the orbital one is 
indicated, and the minima become broader.
Contrary, during $Q \rightarrow A$ transitions in these objects
a rapid decrease in the luminosity of the $L_{\rm ph}$ photons results in
practical disappearance of the H\I\I\ region, and a cool pseudophotosphere
around the hot star is created. Therefore the minima -- eclipses --
are observed at the spectroscopic conjunction and a sudden apparent
change in the period by jump is indicated (Fig. 7, minima just 
before the 1989 outburst).  

\section{Ionization during active phases}

Outbursts of symbiotic stars were originally recognized on the basis 
of their photometric evolution. The most pronounced common feature of 
all active classical symbiotic stars ({\em ClSS}) is an increase in 
the star's brightness, typically by 2-3\,mag, on the time-scale 
of weeks and with the amplitude increasing toward shorter wavelengths. 
The maximum is followed by a gradual decrease to quiescence conditions 
within several months or years. 
However, the light curve profiles and the recurrence of the outbursts are 
very different from object to object. In some cases active phases are 
characterized by multiple maxima (AG\,Dra, Z\,And), in cases of AX\,Per 
and AR\,Pav the optical light during activity waves as a function 
of the orbital phase and the case of CH\,Cyg has no counterpart among 
other symbiotics --  its active phases produce variations in the optical 
continuum up to 6\,mag with superposed changes on the time-scale 
of weeks/hours to seconds. Finally, BF\,Cyg displays three different 
types of outbursts (a symbiotic nova eruption, that observed in 
{\em ClSS} and flares) in its historical light curve [53]. 

Further common observational property of the {\em ClSS} outbursts is
a mass-outflow from the active star in the form of an optically thick 
expanding shell or a fast stellar wind detected spectroscopically 
by profiles of the P-Cygni type and/or by broadening 
of emission lines. In the radio and for the nearest objects 
(e.g. CH\,Cyg, R\,Aqr) the bipolar shaped ejecta (jets) were detected. 
Information about the mass-outflow are, however, sensitive to timing of 
spectroscopic observations (e.g. signatures of the mass ejection are 
better seen at the beginning of outbursts) and the ionization structure 
of the symbiotic nebula (profiles of the highly ionized elements can 
still be rather narrow as they originate at the vicinity of 
the ionizing source, where the material is accelerated to a low velocity).
In addition, the resulting picture depends on the inclination angle. 

To recognize the nature of the contributing lights during outbursts 
we use here a method of analyzing their SED in the 0.12-5.0\,$\mu$ 
and distinguishing between eclipsing and non-eclip\-sing systems. 
This allow us to quantify individual components of radiation and 
to point basic properties of the mass ejection during active phases 
of {\em ClSS}. As an example, we introduce a simple H$\alpha$ method 
to estimate the mass-loss rate during active phases of CH\,Cyg. 
%
%
%
\begin{figure*}[t]
  \centering
  \centerline{\hbox{
  \psfig{figure=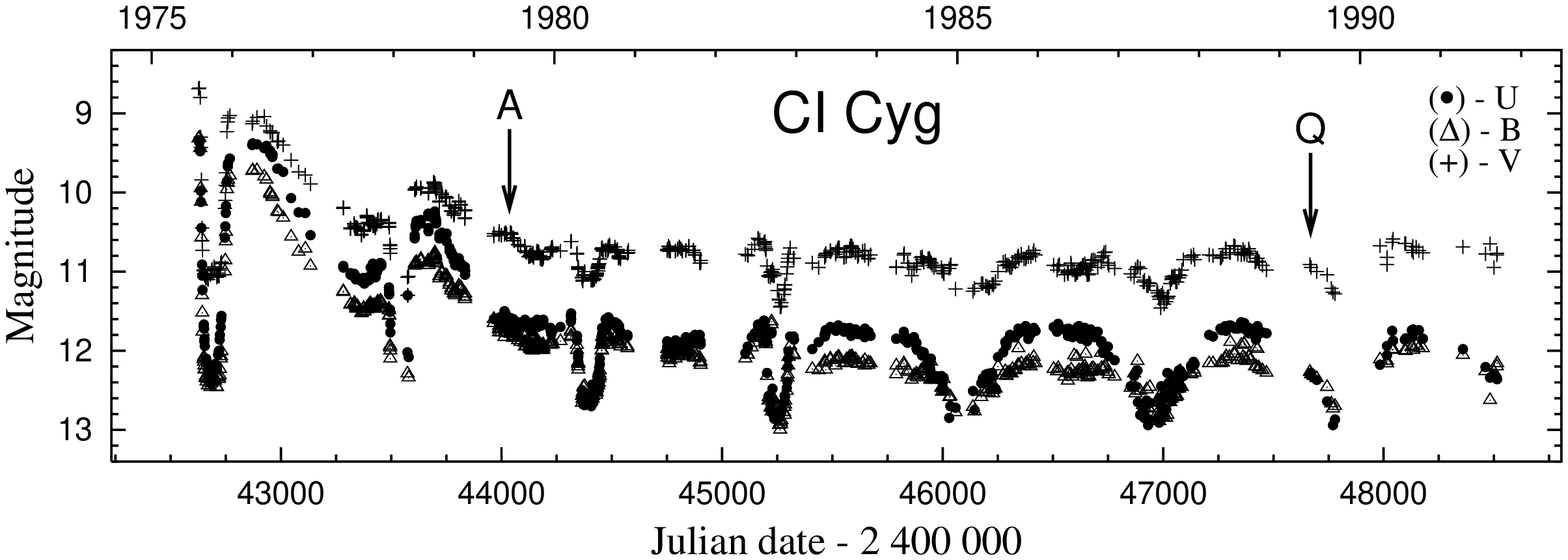,width=17cm,angle=-90}}}
\vspace*{0.25cm}
  \centerline{\hbox{
  \psfig{figure=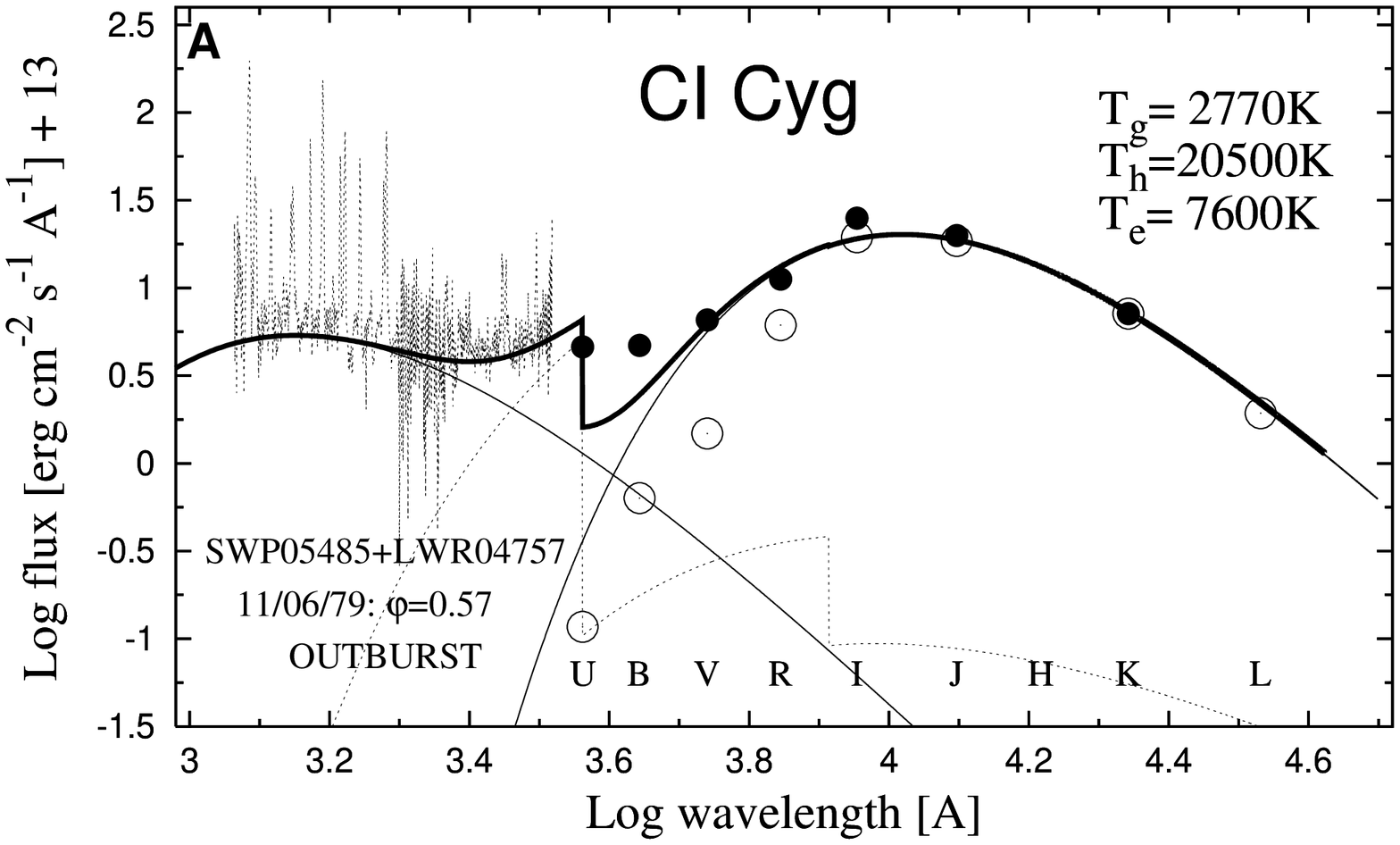,width=8cm,angle=-90}
\hspace*{0.15cm}
  \psfig{figure=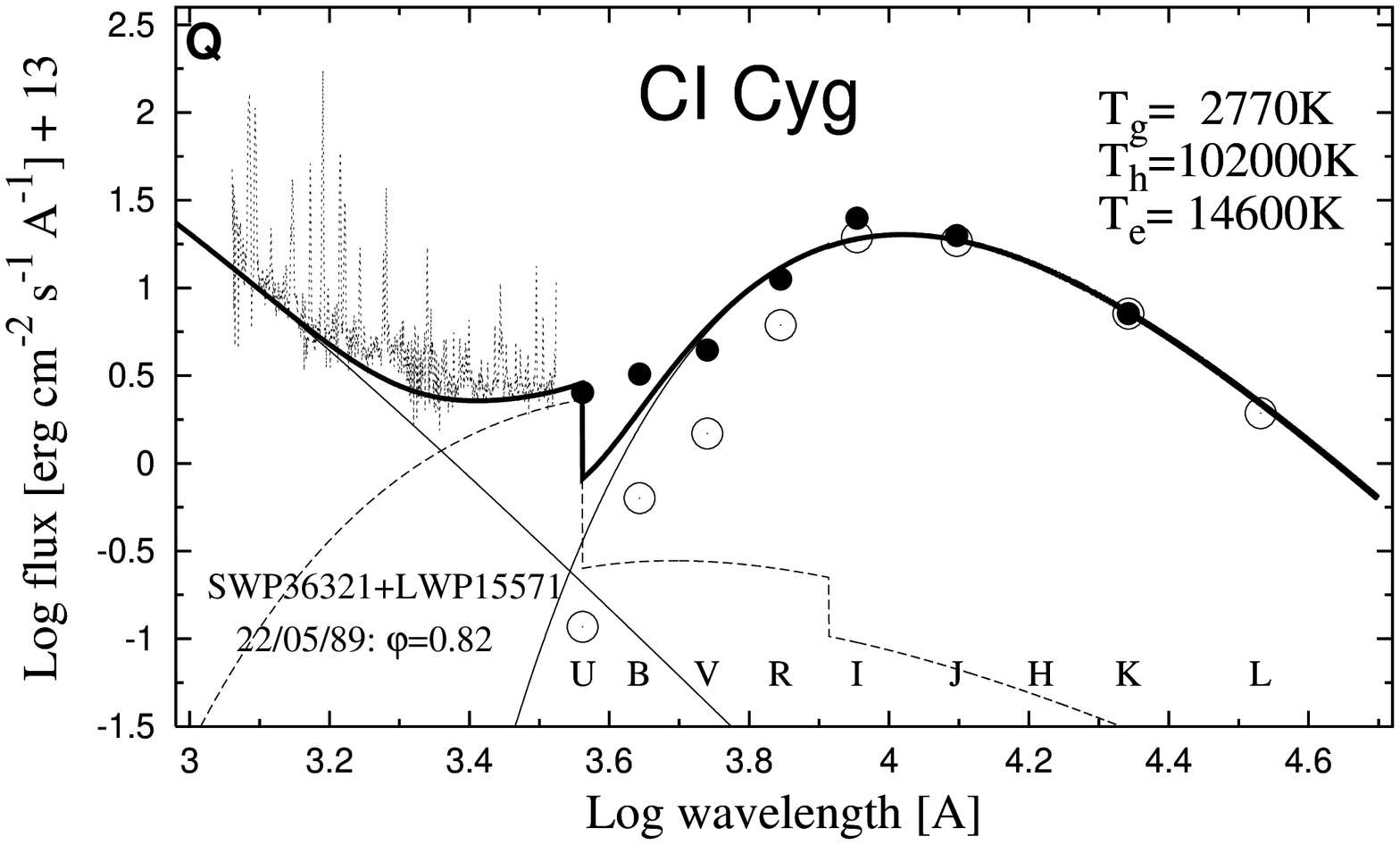,width=8cm,angle=-90}}}
\caption{
Top: The light curve of CI\,Cyg covering the period from its outbursts 
in 1975 to 1993. From about 1984 the system converted to a quiescent 
phase, which is characterized by the wave-like orbitally-related 
variation. The data were summarized from the literature. 
Bottom: The SED during the active phase (left) and quiescence 
(right). This demonstrates a strong change in the ionization structure 
of the CI\,Cyg nebula between both phases (see the text). 
}
\end{figure*}
%
%
%
\begin{figure*}[t]
  \centering
  \centerline{\hbox{
  \psfig{figure=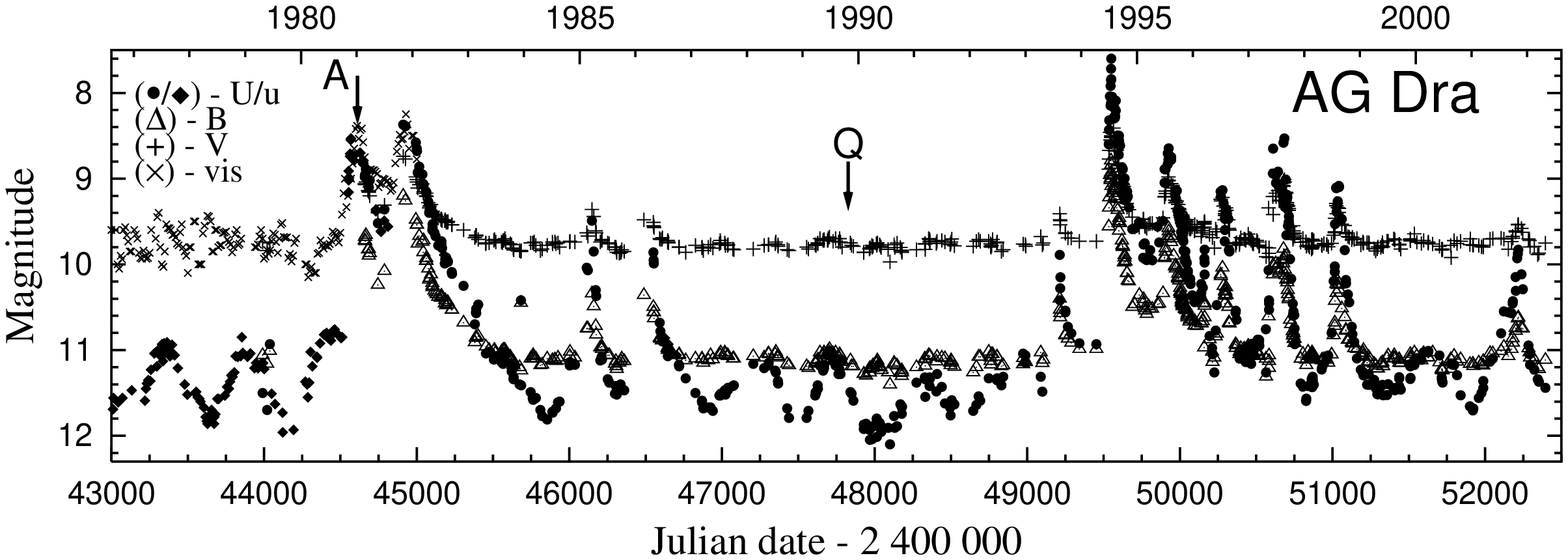,width=17cm,angle=-90}}}
\vspace*{0.25cm}
  \centerline{\hbox{
  \psfig{figure=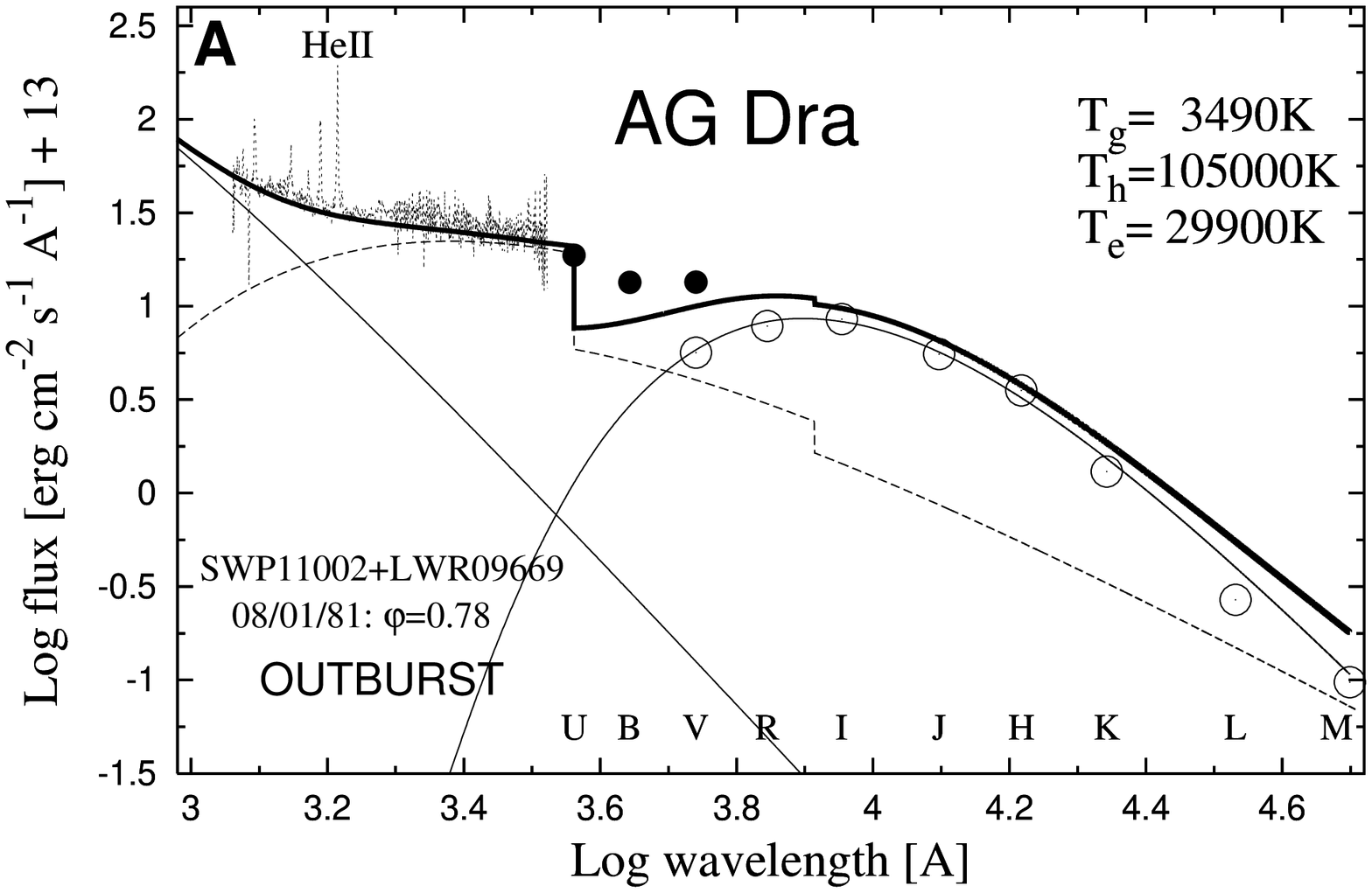,width=8cm,angle=-90}
\hspace*{0.15cm}
  \psfig{figure=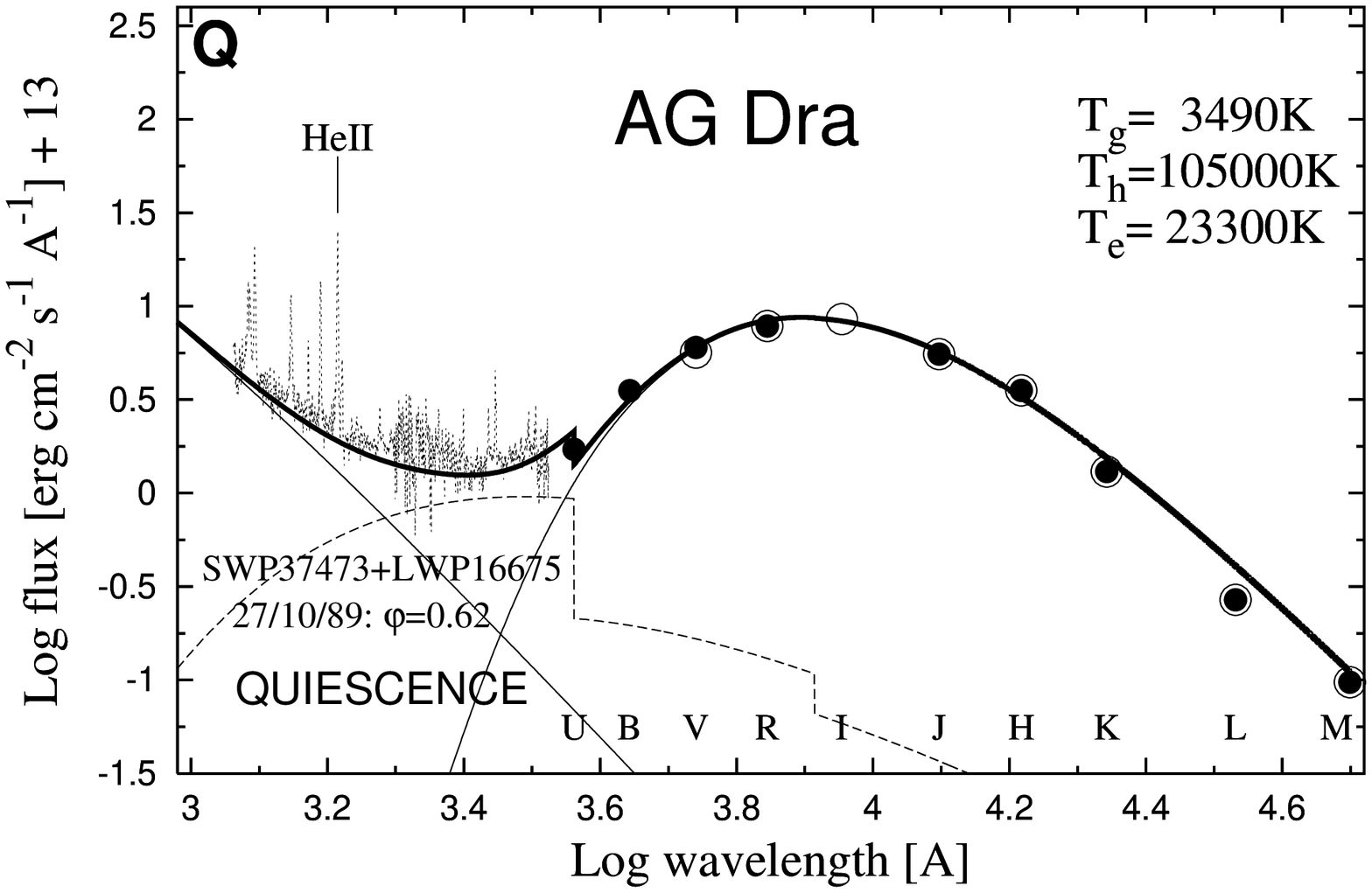,width=8cm,angle=-90}}}
\caption{
Top: The light curve of AG\,Dra covering the period from 1976.7 
to 2002.5. Quiescent phases (to 1981 and 1987-1994.5) are 
characterized by the orbitally-related variation, while the active 
phases by multiple eruptions (1981-82, 1985-86 and 1994-99). 
In both cases the variability is most pronounced in the $U$ 
band. The data were summarized from the literature. 
Bottom: The SED during the 1981 active phase (left) and quiescence 
(right) of the symbiotic binary AG\,Dra. This shows a significant 
contribution of the nebular component of radiation during both phases 
of AG\,Dra.
}
\end{figure*}

\subsection{Eclipsing system CI\,Cyg}

CI\, Cyg is well known eclipsing system with an orbital period of 
855 days. A detailed study of CI\,Cyg made by [24] 
suggests that this system contains an M5\,II giant 
($M_{\rm g} \sim 1.5\,{\rm M_{\odot}}$) 
%
and a hot ionizing source 
($M_{\rm h} \sim 0.5\,{\rm M_{\odot}}$,
%
%
$T_{\rm h} \sim 10^{5}$\,K), 
with a high inclination of the orbit. 
%
%
Fig. 11 shows evolution of the light curve from its major eruption in 
1975. It provides striking constraints on the structure and nature of 
the source of the optical light. The narrow minima during the first 
four cycles from the maximum suggest the source of the optical continuum 
to be formed in a small volume (in the binary dimensions) centered on 
the hot star. This implies that the nebular contribution was negligible 
at the maximum. Contrary to this behaviour, during subsequent cycles, 
the broad minima in the continuum developed. 
The near-UV/optical spectrum was dominated by the nebular emission and 
the lines of highly ionized elements [24]. 

Our fit of the CI\,Cyg SED confirms these results (Fig. 11). 
In addition, such approach allow us to quantify individual components 
of radiation and to determine the corresponding parameters. 
During the active phase a warm blackbody component of radiation 
($T_{\rm h}$ = 20\,500\,K) with superposed highly ionized lines 
dominated the ultraviolet (bottom left panel of Fig. 11). The fit 
suggests the presence of rather faint nebular emission of a very 
low electron temperature ($T_{\rm e}$ = 7\,600\,K). Both components
are subject to eclipse and thus occupy a small region around 
the central star. 
During transition to a quiescent phase (from about 1984) 
the SED in the UV converted to that typical for quiescent phases 
(bottom right panel of Fig. 11, Fig. 1). This reflects a strong 
change in the ionization structure, which is also connected with 
indication of an apparent change of the orbital period as discussed 
in Sect. 3 (see [48] for more details). 
The hot star has a temperature $T_{\rm h} \sim $100\,000\,K and 
the nebula occupies a large volume in the binary. 

\subsection{Non-eclipsing system AG\,Dra}

The binary of AG\,Dra has an orbital period of 550 days (e.g.: [15]) 
and consists of a K-type giant [35] with a mass of 
$M_{\rm g} \sim 1.5\,{\rm M_{\odot}}$ 
      and probably a hot white dwarf of 
$M_{\rm h} \sim 0.5\,{\rm M_{\odot}}$ 
      and 
$T_{\rm h} \sim 0.9-1.3\times 10^{5}$\,K 
embedded in a dense nebula [17,32,43]. 
There are no signs neither in the optical nor far-UV regions of eclipses.
Based on spectropolarimetric observations [43] derived the orbital 
inclination $i=105 - 140^{\circ}$.

The system undergoes occasional eruptions, during which the star's 
brightness abruptly increases ($\Delta U \sim 2,~\Delta V \sim$1\,mag) 
showing multiple maxima separated approximately by 1 year (Fig. 12). 
During eruptions the hot component develops a fast wind at 
a few $\times 10^{2} - 1\,300$\,km\,s$^{-1}$ [60,61,62]. 
The spectral energy distribution of the continuum shows a strong nebular 
component dominating the UV/U-band region mainly during the activity
(Fig. 5 of [18], Fig. 12 here). The quiescent phase of AG\,Dra is
characterized by a periodic wave-like variation in the optical continuum,
which is more pronounced at short wavelengths ([46], Fig. 12).

A precise fitting of the SED of the AG\,Dra continuum (Fig. 4 and bottom 
panels of Fig. 12, see [49] for more details) clarifies 
following points:

(i) Nature of a very different amplitude of the wave-like variation 
in the $UBV$ bands. As only the nebular emission is the subject to 
the wave-like variability (Sect. 2.2.2, Fig. 4), and its contribution 
in the optical is very small with respect to that from the giant 
photosphere, the amplitude in $B$ and $V$ is therefore very small 
with respect to that in $U$. 

(ii) Nature of the UV/optical continuum during active phases. 
The SEDs show that the nebular component of radiation is present 
during both the quiescent and the active phases. In quiescence it 
dominates the near-UV with electron temperature 
$T_{\rm e} \approx\, $20\,000\,K, but in outbursts it represents 
the strongest component between about 1\,500\,\AA\ and the $V$ band 
with $T_{\rm e} \approx\, $30\,000\,K. Also the stellar 
component of the hot star radiation increased considerably by a factor 
of about 10 during the 1981-82 active phase. This point is important 
for comparison to the eclipsing systems. 

\subsection{A mechanism of outbursts}

A strong nebular component of radiation develops immediately 
at the beginning of each outburst of AG\,Dra (Fig. 12, [57]). 
This is in strong contrast to eclipsing systems of BF\,Cyg (Fig. 7) 
and CI\,Cyg (Fig. 11), in which the hot stellar component is replaced 
by a warm, $\sim$20\,000\,K, blackbody radiation at the beginning 
of their outbursts. 
On the other hand, a high-velocity mass-outflow represents 
a common property of outbursts in all systems. 
%

These properties (a mass outflow, but the very different nature 
of the UV continuum) can be interpreted as an inclination effect 
if the outburst's mechanism of the mass ejection is of the same 
nature in both types of systems. \\
{\em Eclipsing systems:} The simultaneous presence of a cool 
pseudophotosphere with the superposed highly ionized lines 
in the ultraviolet suggests an optically thick disk-like torus 
encompassing the hot star to be coincident with the orbital plane. 
Such the formation in the system obscures a fraction of the hot 
star radiation and redistributes it into longer wavelengths. 
The rest part of the hot star flux ionizes the regions above/below 
the disk and thus gives rise to highly ionized lines. The disk is 
subject to eclipses, so its dimension is limited in maximum by 
the radius of the eclipsing cool giant. During transition to 
a quiescent phase it has to dilute or, at least, become to be optically 
thin, because the nebular emission dominates the near-UV/optical region 
(Fig. 11, see also [55]) and occupies a large volume in the binary. 
As a result the wave-like variation in the light curve 
develops (Fig. 11 top, also Figs. 3 and 4 of [55]). \\
{\em Non-eclipsing systems:} The system of AG\,Dra is seen more 
from its pole. Thus if the geometry of the material, which develops 
during outburst, is comparable to the eclipsing systems, the central
ionizing source can be directly seen. As a result both a hot
luminous blackbody and a strong nebular component of radiation 
are observed. The observed increase of the hot star luminosity 
at very high temperature produces a surplus of ionizing photons 
and thus leads to an additional extension of the H\I\I\ zone 
(the parameter $X$ increases, Sect. 2.2.2). In the case of 
an {\em open} ionized 
region ($X > 10$ for AG\,Dra parameters, cf. Fig. 3) a fraction 
of the ionizing photons escapes the system. Therefore an injection 
of new particles (emitters) into such the zone, for example in a form 
of the hot star wind, will produce an extra flux of the nebular nature. 
In the case of very fast wind one can expect an additional ionization 
due to collisions that explains the observed very high electron 
temperature during outbursts. 

Creation of a disk-like structure around the hot star during outbursts 
could be qualitatively understood as a results of an expansion of 
the accretor's shell due to its fast rotation. In this case the hot 
components in symbiotic stars should rotate asynchronously with respect 
to the orbital motion. Contrary, in the case 
of a synchronous rotation, the ejected material will be spread into 
the space rather in form of a shell. This could be the case of BF\,Cyg, 
where only a blackbody component without any emission lines 
was observed in ultraviolet during its 1989 outburst (Fig. 7). 
The above mentioned effect of inclination can be additionally 
complicated by different properties of the mass outflow in different 
objects -- mainly the rate, at which the material is being 
ejected -- that may regulate its optical properties. 

However, this suggestion has to be confirmed by new multi-frequency 
observations of symbiotic stars during outbursts and theoretical 
modeling of the mass-loss to understand better physical processes 
responsible for outbursts of classical symbiotic systems. 
In the following section we propose a simple method to estimate 
the mass-loss rate from the H${\alpha}$ emission. 
%
%
\begin{figure*}[t]
  \centering
  \centerline{
\hbox{\psfig{figure=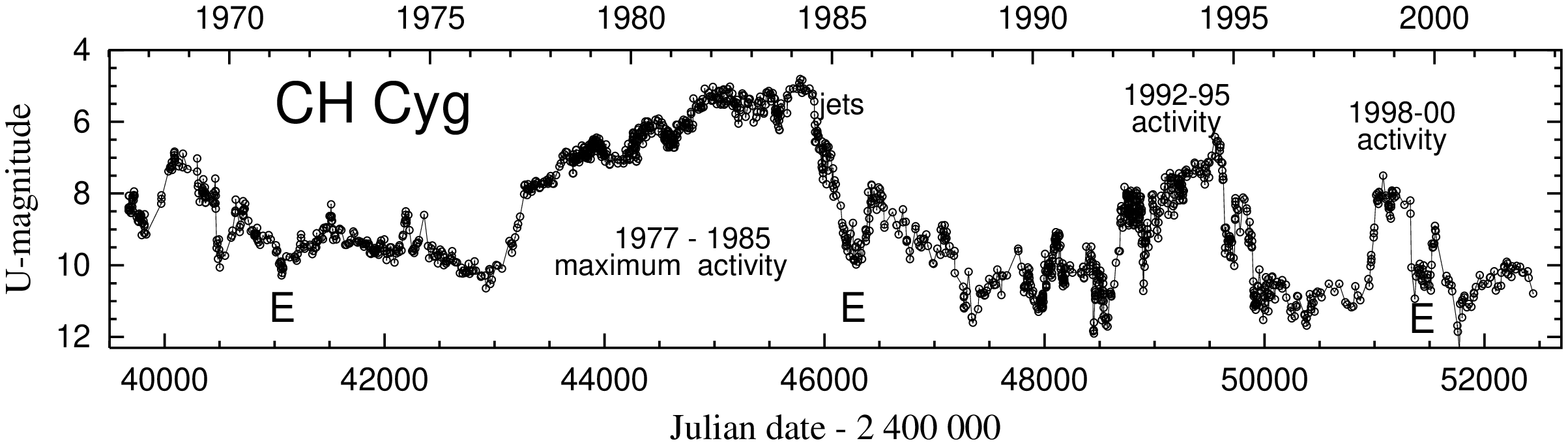,width=17cm,angle=-90}}}
\vspace*{0.25cm}
  \centering
  \centerline{
\hbox{
  \psfig{figure=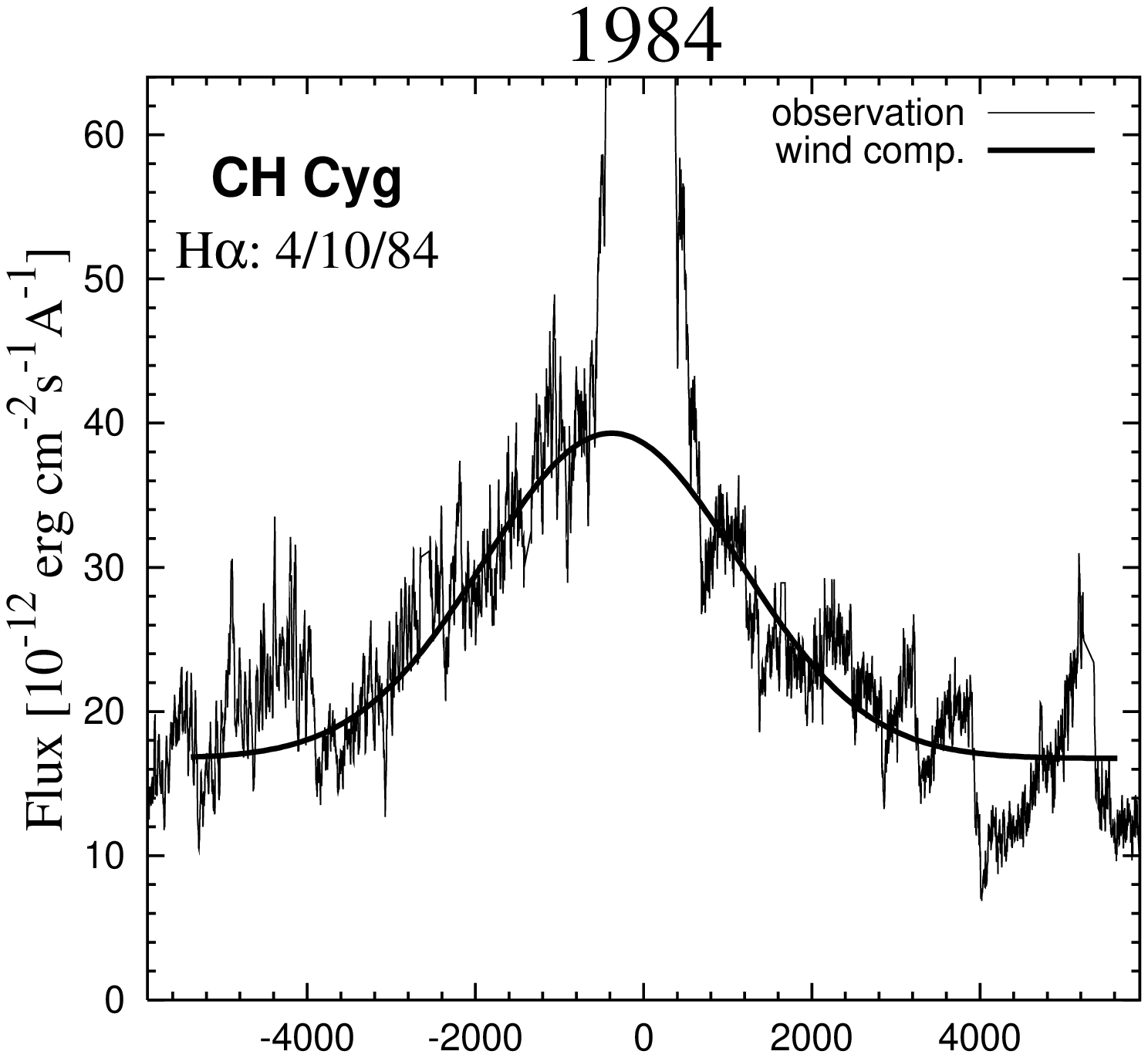,width=4.538cm,angle=-90}
\hspace*{0.5cm}
  \psfig{figure=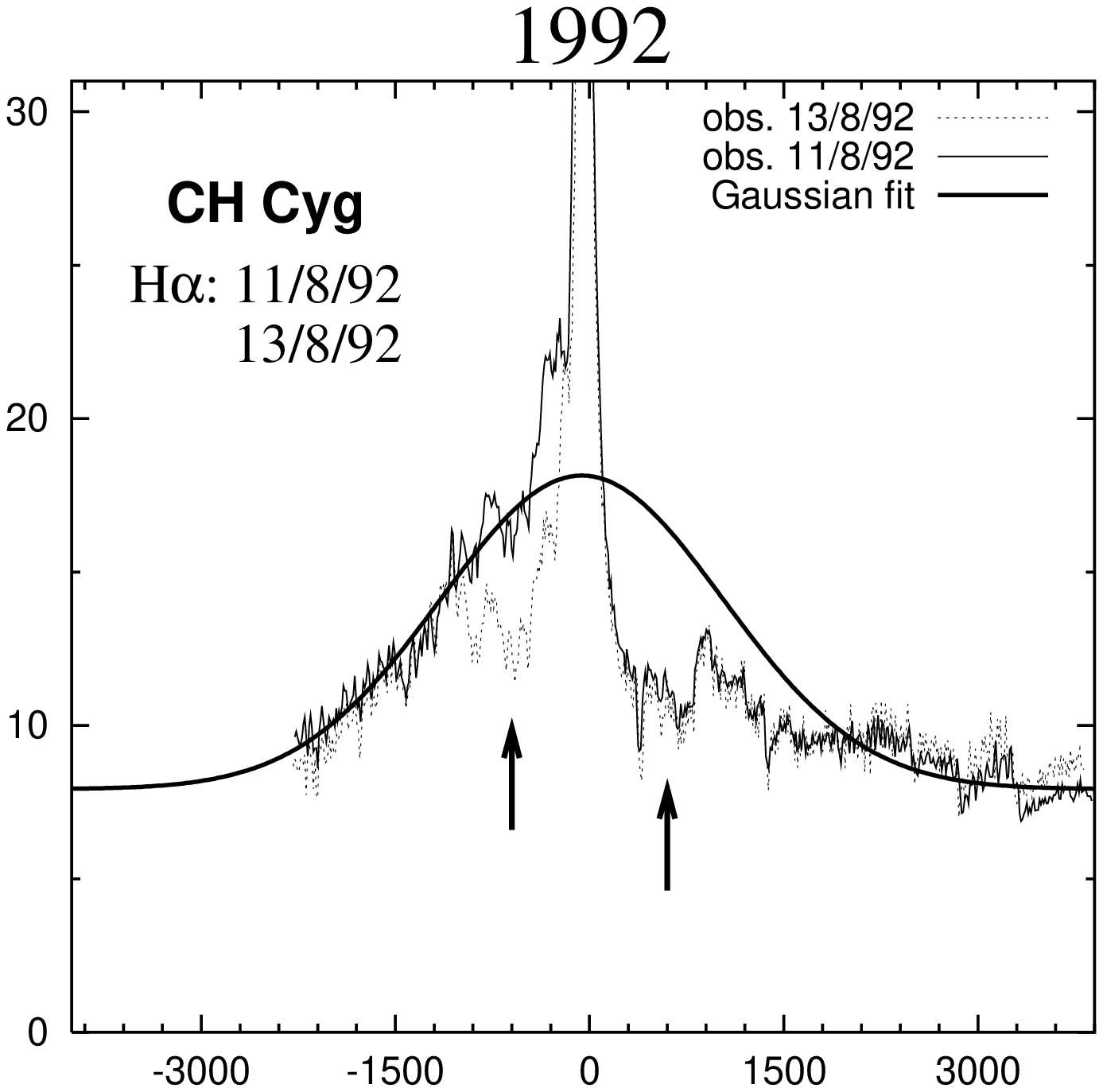,width=4.26cm,angle=-90}
\hspace*{0.5cm}
  \psfig{figure=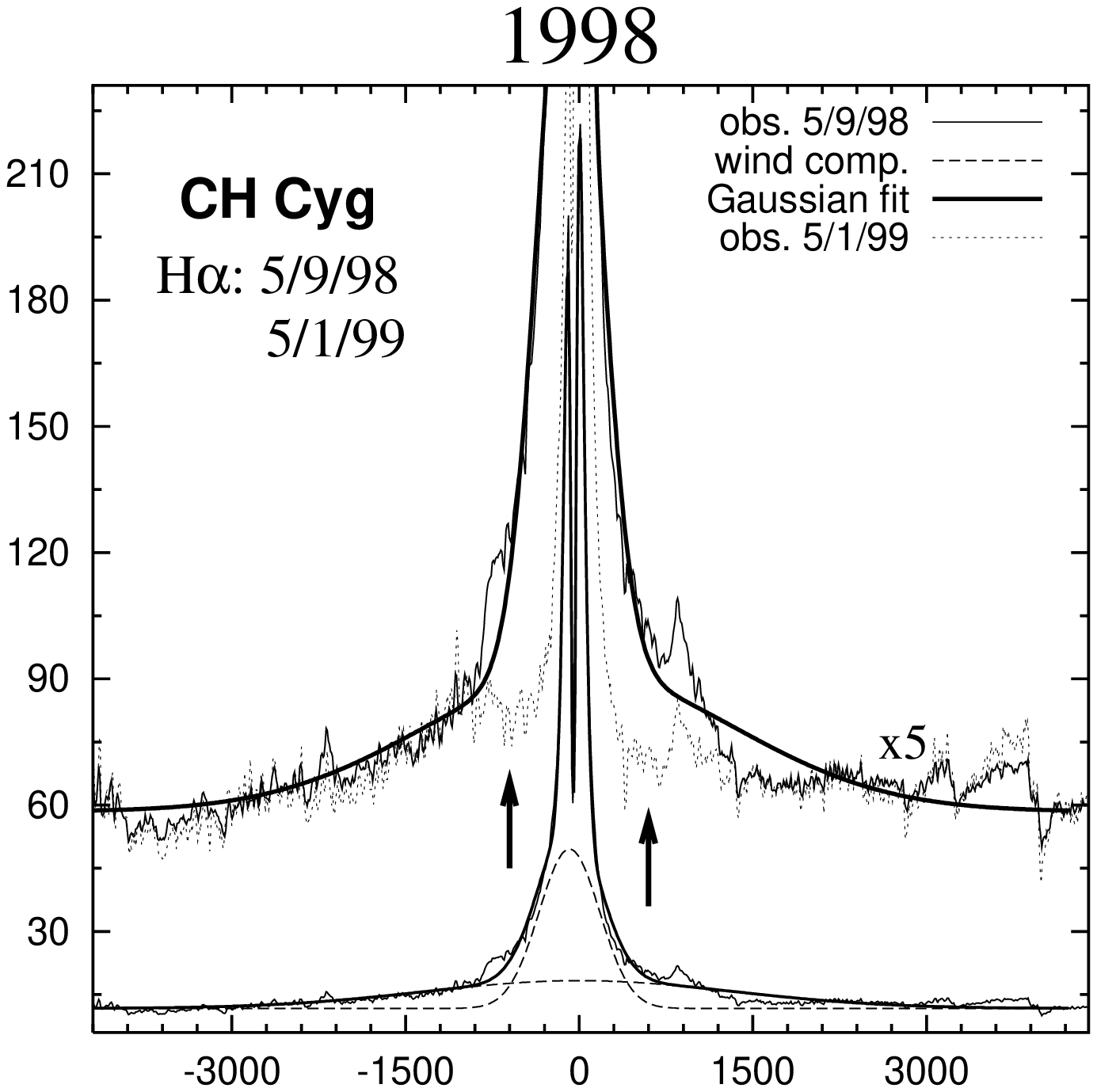,width=4.26cm,angle=-90}}}
\vspace*{0.25cm}
  \centering
  \centerline{
\hbox{
  \psfig{figure=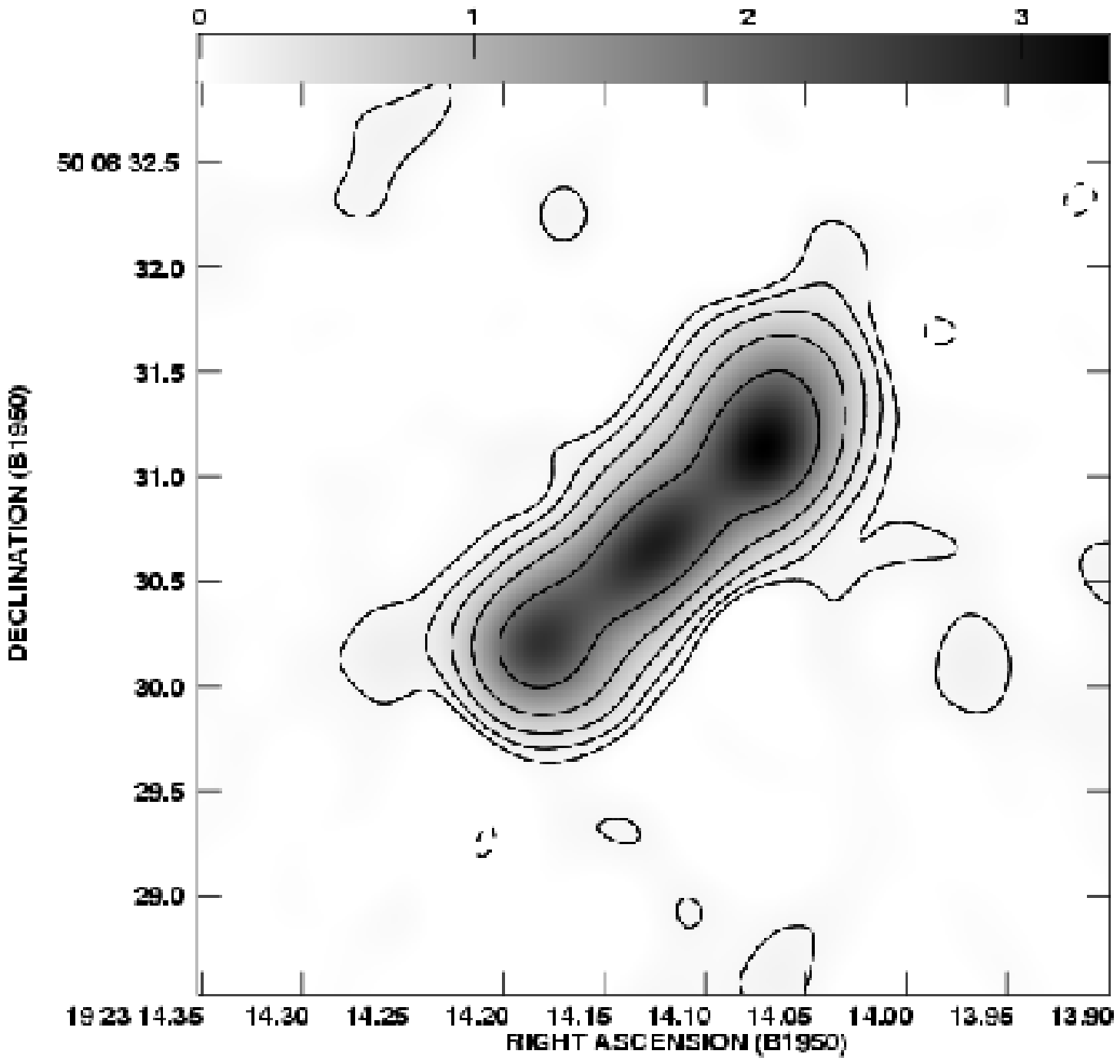,width=4.08cm,angle=-90}
\hspace*{0.8cm}
  \psfig{figure=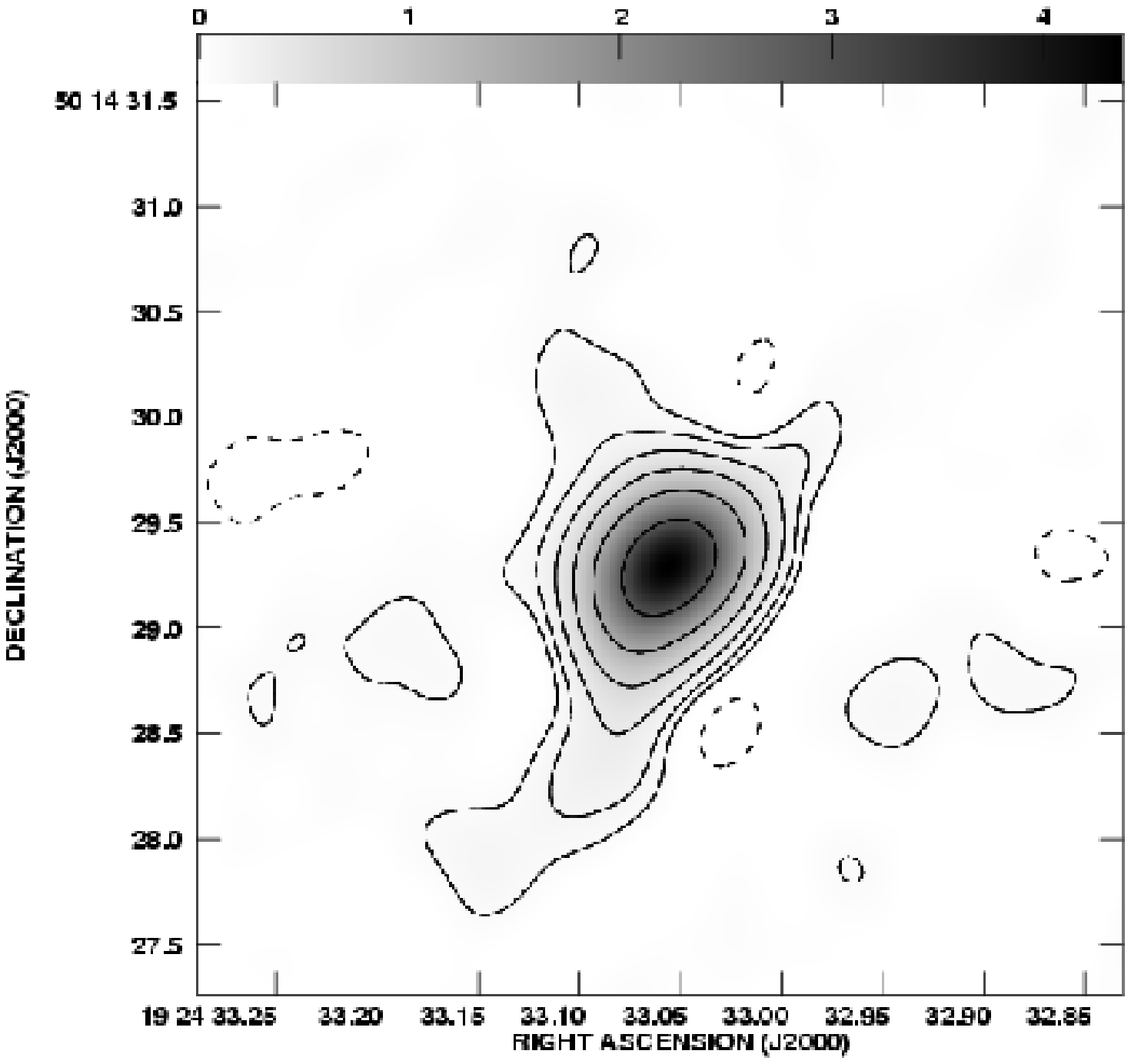,width=4.08cm,angle=-90}
\hspace*{0.8cm}
  \psfig{figure=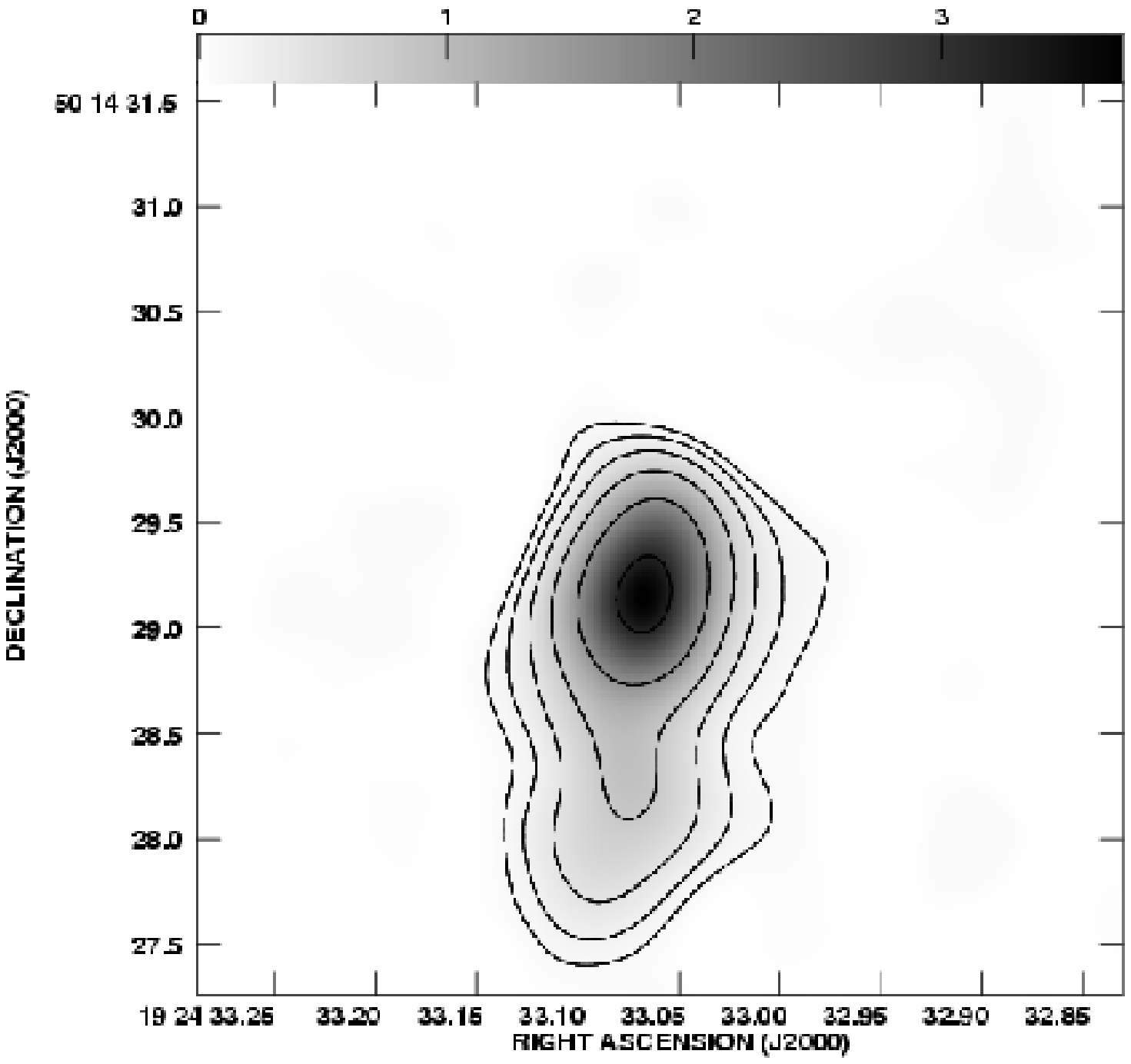,width=4.08cm,angle=-90}}}
\caption[]{
Top: The U-light curve of CH\,Cyg covering the period from 1967 to 2002.5. 
Middle: Hydrogen profiles during three active periods -- in 1984 terminal 
velocities $v_{\infty} \approx 3\,500\,\rm km\,s^{-1}$. In the 1992-95 
active phase $v_{\infty} \approx 3\,000\,\rm km\,s^{-1}$, and absorptions 
on both sides of the line profile at $\pm$(400 -- 600)\,km\,s$^{-1}$ 
appeared (marked by arrows). In the recent, 1998-00, activity basic 
signatures of the profiles were similar to those during the 1992-95. 
Solid thick lines represent the resulting fit by Gaussian functions. 
Broken lines are the broad components of the profile and denoted in 
panels as 'wind components'. 
Bottom: The remnants of the mass loss during activity after 1984.6 as 
detected by the VLA 5GHz radio maps on 1986, 1995 and 1999 (from left 
to right) [7].
}
\end{figure*}

\subsection{Mass loss in the symbiotic system CH\,Cyg}

CH\,Cygni is a mysterious symbiotic star. It started a symbiotic 
activity in 1963. Prior to this, CH\,Cyg had been observed 
as a semi-regular variable (e.g.: [33]). Subsequently, the symbiotic 
phenomenon was observed during 1967-86, 
1977-86, 1992-95 and 1998-00 (Fig. 13). 
Recent studies suggest that CH\,Cyg is an eclipsing triple-star system
consisting of the inner, 756-day period binary (the symbiotic pair, which
is responsible for the observed activity) and another cool giant 
revolving around it on the long, 14.5-year period orbit ([20,51,54], 
original suggestion of a triple-star model was set by [19]). 
The distance to CH\,Cyg was determined on the basis of Hipparcos 
measurements as $270\,\pm\,66$\,pc (e.g.: [63]). 

Active phases of CH\,Cyg after 1984.6 were characterized by pronounced 
signatures of a high-velocity mass outflow with terminal velocities 
of 2\,500 -- 3\,500\,km\,s$^{-1}$. This and the vicinity of CH\,Cyg 
allowed us to obtain its spatially resolved images. For the first 
time [58] detected bipolar jets by the VLA radio observation. 
Further observations of the CH\,Cyg ejecta revealed a non-thermal 
emission at its outer parts and a precession with a period of 
$\approx$ 20\,years [7,8]. A detail structure given by the 
ground-based observations and the HST was studied by [6,9]. Figure 13 
shows some examples of the high-velocity features of CH\,Cyg. 

The rate, at which the material is ejected by the central star, 
represents a very important parameter to understand 
a balance between the input and output of the energy during 
outbursts, ionization structure of the symbiotic nebulae, 
a mechanism of the mass-loss, etc. 
For symbiotic binaries an approximative approach to estimate 
the mass-loss rate for EG\,And, AG\,Peg and AG\,Dra was used by
[61,64,65]. In all cases the authors assumed an optically thin 
wind at a constant velocity ($v=v_{\infty}$).
In the following section we introduce a simple H$\alpha$ method     
to investigate the mass-loss rate during active phases of CH\,Cyg 
as recently suggested by [56]. 

\subsubsection{Mass loss from the H${\alpha}$ emission}

In the recombination process, the transition from the third to the second
level of hydrogen has a high probability of producing the H$\alpha$ 
line in emission. In the case of an ionized stellar wind, the H$\alpha$
profile contains information on its emissivity and velocity 
distribution. Thus, having a theoretical H$\alpha$ luminosity and
some knowledge of the velocity field of the outflow, one can derive
the mass-loss rate, $\dot M$, from the measured strength of 
the H$\alpha$ emission. 
Here we use a simplified approach assuming that the wind is
(i)   spherically symmetric with a steady $\dot M$,
(ii)  fully ionized and completely optically thin in H$\alpha$ from 
      a radius $r_{\rm min}$ above the ionizing source and 
(iii) isothermal at the same temperature as the stellar photosphere. 
The total line luminosity, $L(\rm H\alpha$), is related to the line 
emissivity of the wind, $\varepsilon_{\alpha} n_{\rm e}n^{+}$, by
\begin{equation}
  L({\rm H}\alpha) = 4\pi d^{2} F({\rm H}\alpha)
                  = \varepsilon_{\alpha}\!\! \int_{V}\!\!
                             n_{\rm e}n^{+}(r)[1-w(r)] \,{\rm d}V,
\end{equation}
where $d$ is the distance to the system, $F({\rm H}\alpha)$ represents
the observed flux in H$\alpha$ ($\rm erg\,cm^{-2}\,s^{-1}$), 
$\varepsilon_{\alpha}$ is the volume emission coefficient in H$\alpha$, 
$n_{\rm e}$ and $n^{+}$ are number densities of electrons and 
ions (protons) and the factor $w(r)$ is the fraction of the solid angle 
that is covered by the photosphere (so called geometrical dilution 
factor). Further assumption of a constant value of 
$\varepsilon_{\alpha} = 3.56 \times 10^{-25} \rm erg\,cm^{3}\,s^{-1}$ 
throughout the ionized medium at the electron temperature of 
$T_{\rm e}$ = 10$^{4}$\,K simplifies considerably the approach. 
For a completely ionized medium, ($n_{\rm e}\simeq n^{+}$), 
the particle density $n(r)$ in the wind can be expressed in terms
of the mass-loss rate and the velocity law via the mass continuity
equation as
\begin{equation}
 n(r) = \dot M/4\pi r^{2}\mu m_{\rm H} v(r),
\end{equation}
where $\mu$ is the mean molecular weight, $m_{\rm H}$ is the mass of
the hydrogen atom and $v(r)$ is the velocity distribution in the hot
star wind, approximation of which is given by Eq. 5. According to 
these assumptions, [56] derived an expression for the H$\alpha$ 
luminosity as 
\begin{equation}
  L({\rm H}\alpha) = \frac{\varepsilon_{\alpha}}{4\pi(\mu m_{\rm H})^2}
	  \Big(\frac{\dot M}{v_{\infty}}\Big)^{2}\frac{1}{R_{\star}}
	  \times I_{\rm w},
\end{equation}
where the integral $I_{\rm w}$ is a function of the $\gamma$ and 
$r_{\rm min}$ parameters (see [56] for more detail). Then comparing 
the observed H$\alpha$ luminosity from the wind with Eq. 18, one 
can obtain the mass-loss rate. 
[56] estimated the $\gamma$ parameter by comparing 
a synthetic-line profile to the observed one. They obtained average 
values of $\gamma$ = 1.4 and 1.9 for 1984.6-92 and 1998-00 active phases, 
respectively. For other input quantities, the radius of the active star 
in CH\,Cyg 
$R_{\star} = 5\,\rm R_{\odot}$, 
$r_{\rm min} = 8-9\,\rm R_{\odot}$ 
and the outer radius of 200\,R$_{\odot}$ (= a finite limit, at which 
the wind contributions are negligible), they determined the mass-loss 
rate from the active star 
\begin{eqnarray}
 \dot M_{1984}\,=\,(4.4\pm 1.5)\times 10^{-6}\rm M_{\odot}\, yr^{-1}~ 
\nonumber  \\
  \dot M_{1998}\,=\,(1.8\pm 0.7)\times 10^{-6}\rm M_{\odot}\, yr^{-1}.
\end{eqnarray}
These values correspond to the measured H${\alpha}$ fluxes converted 
to the scale of luminosities for the distance of 270\,pc. It is 
important that our $\dot M$ estimates agree well with mass-loss rates 
derived from radio observations [52,58]. The radio mass-loss rates are 
less model dependent than the H$\alpha$ method. In the former case 
we observe the ejected material at very large distances from the star 
(from a few 10's of A.U. to $\approx$1\,000\,A.U.), where a constant 
expansion velocity can be reliably assumed. On the other hand, in 
the latter approach we are dealing with regions at distances of a few 
R$_{\odot}$ to, at maximum, a few hundreds of R$_{\odot}$, the emissivity 
of which critically depends on the velocity law, whose structure, however, 
has to be assumed. 

\section{Conclusions} 

Symbiotic stars represent very suitable space laboratories for 
stu\-dying processes of ionization and recombination. This paper 
reviewed some recently investigated effects of these processes. 
It was found that the ionization structure of symbiotic nebulae 
can be very different and variable during different levels of 
activity and thus can significantly influence our observations. 

During {\em quiescent} phases, when the compact hot star produces 
the energy at approximately the same amount and temperature, 
an effect mimicking the reflection effect takes place. It was 
found that the observed total emission from the symbiotic nebula 
is consistent with that given by a simple ionization model and 
the variation in the emission measure is fully responsible for 
variation observed in the light curves. To explain the orbitally-related 
variation, it is assumed that the nebula is partially optically thick 
and of a non-symmetrical shape. Future work should thus include 
investigation of the optical properties of symbiotic nebulae. 
First, under conditions of a simple ionization model, with further 
implementation of effects of the wind accretion onto 
the compact object. The latter is closely connected with 2D and 3D 
hydrodynamical calculations of gas flow in symbiotic stars to model 
a more realistic density distribution (e.g.: [16,67]). 
Also this is important to continue the long-term photometric monitoring 
of well studied symbiotic stars to obtain information on the geometry 
and location of the optically thick portion of the symbiotic nebula 
in the system. An extension of such the monitoring programme to 
the near-IR region is desirable to search for a double-wave variability 
in the $VRI$ bands to test the possibility of an ellipsoidal 
shape of red giants in symbiotic binaries. 

During {\em transition} periods -- from activity to quiescence -- 
a gradual variation in the luminosity of photons capable ionizing 
hydrogen results in a systematic variation in the structure of the 
H\I\I\ region. As the symbiotic nebulae are not symmetrically placed
around the binary axis, their gradual change in the geometry and 
location in the binary can produce the observed variation in 
the $O-C$ residuals and thus an apparent change in the period.  
This effect is best indicated for eclipsing systems. 

During {\em active} phases the ionization structure of symbiotic 
nebulae is disrupted by the mass flow from the active star. 
The result depends on the mass outflow properties and the binary 
parameters -- mainly the orbital inclination. By disentangling 
the observed SED for the example objects, it was found that 
in the non-eclipsing systems (e.g. AG\,Dra) the nebular component 
of radiation in the continuum considerably strengthened at a high 
electron temperature of about 30\,000\,K. On the other hand, 
eclipsing systems (e.g. CI\,Cyg) display simultaneous presence of 
a cool pseudophotosphere with the superposed highly ionized lines
in the ultraviolet. This suggests an optically thick disk-like torus
encompassing the hot star and being coincident with the orbital plane.
The nebular contribution in the continuum here is very faint 
and corresponds to a low electron temperature of $\sim$8\,000\,K. 
To understand better the mechanism of the mass ejection, 
multi-frequency observations of symbiotic stars during outbursts 
and theoretical modeling of the mass outflow are highly desirable. 
Example of an improved method to estimate the mass-loss rate 
during active phases of CH\,Cyg suggested values of 
$\dot M = (1.8 - 4.4)\times 10^{-6}\rm M_{\odot}\, yr^{-1}$. 
We propose to apply, at least, this approach to other systems 
with significant features of a high-velocity mass outflow. \\

\noindent
{\em Acknowledgments.}
A part of this research has been done within the project 
No. SLA/1039115 of the Alexander von Humboldt foundation during 
the author's visits at the Astronomical Institute in Bamberg, in part 
throughout the bilateral research project between the Royal Society 
of Great Britain and the Slovak Academy of Sciences and also was 
partly supported by the Slovak Academy of Sciences grant No. 1157 
and by the APVT-20-014402 Project.

\end{document}